\documentclass[lettersize,journal]{IEEEtran}
\usepackage{amsmath,amsfonts}
\usepackage{array}
\usepackage[caption=false,font=normalsize,labelfont=sf,textfont=sf]{subfig}
\usepackage{textcomp}
\usepackage{stfloats}
\usepackage{url}
\usepackage{verbatim}
\usepackage{graphicx}
\usepackage{cite}
\usepackage{booktabs}
\usepackage{xcolor}
\usepackage{amssymb}
\usepackage{pifont}
\usepackage{longtable}
\usepackage{makecell}
\usepackage{rotating}
\usepackage{multirow}
\usepackage[normalem]{ulem}
\usepackage[most]{tcolorbox} 
\usepackage{siunitx}
\newcolumntype{L}[1]{>{\raggedright\arraybackslash}p{#1}}

\usepackage[hidelinks]{hyperref}
\usepackage{orcidlink}

\definecolor{chestnut}{rgb}{0.97, 0.51, 0.47}
\def\BibTeX{{\rm B\kern-.05em{\sc i\kern-.025em b}\kern-.08em
    T\kern-.1667em\lower.7ex\hbox{E}\kern-.125emX}}

\begin{document}

\bstctlcite{IEEEexample:BSTcontrol}

\title{Enabling Cloud-Level Accuracy in Edge AI through IoT Data Preprocessing}

\author{Ayg\"{u}n~Varol\,\orcidlink{0000-0002-4029-7676}\textsuperscript{1},
Katarzyna~Ko{\l}odziej\,\orcidlink{0000-0002-2329-107X}\textsuperscript{2},
{\L}ukasz~Sobczak\,\orcidlink{0000-0001-9439-1812}\textsuperscript{2},
Micha{\l}~Romaszewski\,\orcidlink{0000-0002-8227-929X}\textsuperscript{2},
Przemys{\l}aw~G{\l}omb\,\orcidlink{0000-0002-0215-4674}\textsuperscript{2},
Naser~Hossein~Motlagh\,\orcidlink{0000-0001-9923-9879}\textsuperscript{3},
Mirka~Leino\,\orcidlink{0000-0002-0465-4197}\textsuperscript{4},
and~Johanna~Virkki\,\orcidlink{0000-0002-2216-7296}\textsuperscript{1}%
\thanks{\textsuperscript{1}Faculty of Information Technology and Communication Sciences, Tampere University, Tampere, Finland.}
\thanks{\textsuperscript{2}Institute of Theoretical and Applied Informatics, Polish Academy of Sciences, Gliwice, Poland.}
\thanks{\textsuperscript{3}Department of Computer Science, University of Helsinki, Helsinki, Finland.}
\thanks{\textsuperscript{4}Satakunta University of Applied Sciences, Pori, Finland.}
\thanks{Corresponding author: Ayg\"{u}n Varol (e-mail: aygun.varol@tuni.fi).}
}

\maketitle

\begin{abstract}
Large language models (LLMs) offer a natural-language interface for interpreting Internet of Things (IoT) sensor data in smart environments; however, cloud-based deployment introduces latency, privacy, and connectivity concerns. Local LLM deployment can reduce these limitations, but compact edge-deployable models often show weaker numerical reasoning when raw sensor readings are provided directly. This paper investigates whether prompt-side preprocessing can improve the accuracy--latency trade-off of local LLMs for environmental monitoring tasks. We propose a structured prompt construction framework that transforms raw air-quality and thermal-comfort measurements into increasingly enriched textual representations: raw sensor values, threshold-aware descriptions, and compact environmental summary flags. The approach is evaluated using indoor datasets collected with Raspberry Pi nodes and BME680 sensors at Tampere University and outdoor air-quality datasets from Helsinki, Katowice, and Warsaw. We construct a binary LLM query dataset covering air quality, thermal comfort, and joint environmental conditions, and compare five local LLMs and five cloud LLMs across three prompt variants and two inference modes, with and without chain-of-thought prompting. The results show that prompt enrichment substantially improves local-model accuracy. In No-CoT mode, local accuracy increases from 50.9\% to 81.7\% indoors and from 63.7\% to 89.3\% outdoors from the raw-prompt setting to the most enriched prompt. Latency results show that local No-CoT inference is the fastest configuration, with mean latency close to 0.22s per inference, while CoT substantially increases inference time. These findings suggest that lightweight prompt-side preprocessing can narrow the local--cloud performance gap and support practical low-latency IoT analytics for smart environments.

\end{abstract}


\section{Introduction}\label{sec:Introduction}

Smart environments integrate Internet of Things (IoT) devices, sensors, actuators, communication networks, and computing resources to monitor physical conditions and support automated or user-assisted decision-making. By enabling real-time sensing and data-driven control, these environments can improve energy efficiency, resource management, operational cost, occupant comfort, and safety~\cite{de2024smart,varol2026creation}. They are increasingly deployed in smart buildings, indoor workspaces, healthcare facilities, industrial environments, and smart cities, where continuous sensing can support adaptive services and user-centric automation~\cite{yan2025general}. Among these applications, air-quality and thermal-comfort monitoring are particularly important because they directly affect occupant well-being, productivity, and safety. Indoor air quality (IAQ) monitoring commonly involves parameters such as particulate matter, volatile organic compounds, carbon monoxide, carbon dioxide, temperature, and humidity, and IoT-based IAQ systems have been shown to support real-time monitoring, health-risk awareness, and ventilation optimization~\cite{tan2024revolutionizing, motlagh2023digital}.

\begin{figure}
  \centering
  \includegraphics[width=\linewidth]{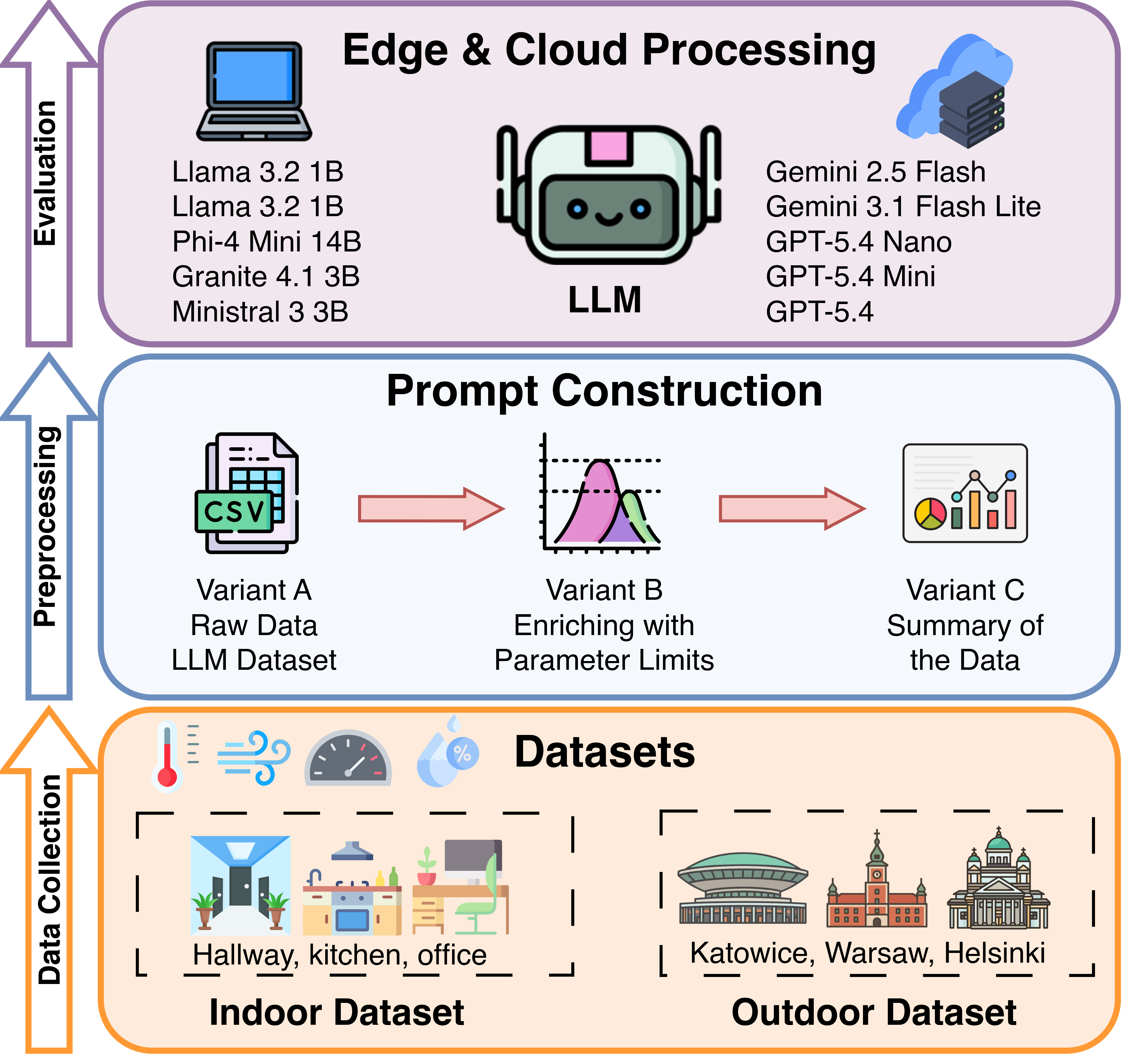}
  \caption{Overview of the proposed IoT data preprocessing and LLM-based environmental analytics framework. Sensor data are collected from indoor and outdoor environments, transformed into progressively enriched prompt representations, and evaluated using local and cloud LLMs.}
  \label{fig:Architecture}
\end{figure}

Recent advances in artificial intelligence (AI) have expanded the capabilities of smart environments beyond simple rule-based automation. Machine learning and deep learning methods have been used for activity recognition, anomaly detection, energy optimization, environmental prediction, and proactive control~\cite{varol2026creation}. However, conventional AI pipelines often require task-specific data collection, feature engineering, model training, and periodic retraining when the sensing environment or decision criteria change. This creates a practical barrier for dynamic smart environments, where sensors, locations, user requirements, and regulatory thresholds may vary over time. Moreover, many traditional AI-based interfaces remain limited in their ability to interpret flexible user queries or explain environmental conditions in a natural, human-centered form. 

Large language models (LLMs) provide a promising alternative interface for IoT analytics since they can interpret natural-language queries, combine heterogeneous contextual information, and generate human-readable responses. In smart environments, this capability can reduce the technical barrier between users and sensor data by allowing occupants, facility managers, or operators to ask questions such as whether indoor air quality is acceptable, whether thermal conditions are comfortable, or whether outdoor conditions are suitable for a given activity. Recent smart-building and smart-space studies show that LLMs can support natural-language interaction, reasoning, action planning, and tool use in IoT-enabled environments~\cite{yan2025general,varol2025performance, varol2026multi}. Therefore, LLMs offer a pathway toward more flexible and user-oriented environmental analytics than fixed dashboards or application-specific classifiers.

Despite this potential, most high-capacity LLMs are deployed on cloud platforms due to their substantial computational and memory requirements. Cloud-based LLMs provide strong analytical capabilities, yet they introduce latency, privacy, connectivity, availability, and sustainability concerns~\cite{semerikov2025llm,singh2025survey}. These concerns are especially relevant in smart-environment and IoT settings, where sensor data may include sensitive information about occupants and where local operation may be required during network disruptions or emergency conditions. Local LLM deployment on edge devices can reduce cloud dependence and improve data locality, yet compact local models typically have weaker numerical reasoning and lower instruction-following robustness than cloud models. Surveys on edge LLMs and mobile edge intelligence emphasize that privacy-sensitive and delay-sensitive applications motivate edge deployment, while also highlighting the resource limitations of on-device inference~\cite{friha2024llm,qu2025mobile}.

Hybrid edge--cloud architectures have emerged as a practical way to balance these trade-offs. By combining local processing with cloud-based reasoning, they can use edge resources for low-latency or privacy-sensitive tasks while reserving cloud models for complex analysis~\cite{friha2024llm,varol2025performance,verma2017survey}. However, local LLM performance in IoT analytics depends not only on model size or deployment location, but also on how numerical sensor measurements are represented. Raw environmental readings can be difficult for compact LLMs to interpret reliably, especially when comparison with regulatory or comfort thresholds is required. This motivates prompt-side preprocessing as a lightweight alternative to fine-tuning.

This paper investigates whether prompt-side preprocessing can improve the accuracy--latency trade-off of local LLMs for environmental IoT analytics. The key idea is to transform raw sensor measurements into progressively enriched textual representations before they are provided to the model. We evaluate three prompt variants: raw sensor- and limit-value prompts, threshold-aware prompts that include limit-comparison descriptions, and status-enriched prompts that add compact environmental summary flags. Fig.~\ref{fig:Architecture} presents the overall architecture, consisting of data collection, prompt-side preprocessing, and local/cloud LLM processing layers.

We study this problem through air-quality and thermal-comfort classification. The evaluation uses indoor datasets collected with Raspberry Pi nodes and BME680 sensors at Tampere University, together with outdoor air-quality datasets from Helsinki, Katowice, and Warsaw. We construct a binary LLM query dataset covering air-quality, thermal-comfort, and joint environmental questions. Local and cloud LLMs are evaluated across three prompt variants and two inference modes: direct answering without chain-of-thought reasoning (No-CoT) and chain-of-thought prompting (CoT). The evaluation reports accuracy and latency to capture classification performance and deployment feasibility. More broadly, this work contributes to Edge AI by showing how prompt-side optimization can help smaller local LLMs achieve useful environmental-monitoring performance with reduced computational overhead.

The main contributions of this paper are as follows:

\begin{itemize}
    \item \textbf{Prompt-side preprocessing framework for local LLMs:}
    We propose a structured prompt construction method that transforms raw IoT sensor readings into threshold-aware and status-enriched textual representations for environmental decision-making.

    \item \textbf{LLM query dataset for environmental monitoring:}
    We construct a binary question-answer dataset covering air quality, thermal comfort, and joint environmental conditions for indoor and outdoor sensor records.

    \item \textbf{Local--cloud evaluation of LLM-based IoT analytics:}
    We evaluate five local LLMs and five cloud LLMs across three prompt variants and two inference modes, enabling a controlled comparison of prompt enrichment, deployment setting, and reasoning strategy.

    \item \textbf{Accuracy and latency analysis:}
    We jointly analyze classification accuracy, output-format reliability, and response latency, showing that enriched No-CoT prompts provide the most practical configuration for low-latency local environmental classification.
\end{itemize}

The remainder of this paper is organized as follows. Section~\ref{sec:RelatedWork} reviews related work on LLMs in smart environments, sensor interpretation, edge/cloud deployment, prompting strategies, and air-quality analytics. Section~\ref{sec:Method} presents the proposed methodology and prompt construction variants. Section~\ref{sec:Dataset} presents the indoor and outdoor datasets and the LLM query dataset construction. Section~\ref{sec:Experimental} describes the experimental setup, inference modes, and evaluated models. Section~\ref{sec:Results} reports aggregate accuracy and latency trends, and presents detailed model- and location-level results. Section~\ref{sec:Discussion} discusses the implications and limitations of the findings, and Section~\ref{sec:Conclusion} concludes the paper.

\section{Related Work}\label{sec:RelatedWork}

\textbf{LLMs in Smart Environments and IoT Agents}: A growing body of literature has explored LLMs as natural-language interfaces and orchestration layers for smart environments and IoT systems. The study in~\cite{king2023get} showed that GPT-based models can infer user intent from abstract smart-home commands, and translate them into machine-readable actions for controlling connected devices. ChatIoT~\cite{gao2024chatiot} extends this direction by generating trigger-action programs from natural-language requests, using context-aware compressive prompting to improve token efficiency and generation accuracy in home automation scenarios. In smart buildings, BuildingSage~\cite{dedeoglu2024buildingsage} proposes a privacy-preserving AI copilot that enables users to interact with building data through natural language while executing generated code locally and exposing only metadata to remote LLMs. The study in~\cite{saleh2025follow} further illustrates how AI agents can support personalized interaction with smart environments by coordinating user preferences, sensor data, environmental controls, and device--edge--cloud resources.

LLM-based agents have also been proposed for more complex IoT coordination. LLMind~\cite{cui2024llmind} introduces a task-oriented IoT framework in which an LLM coordinates domain-specific AI modules and IoT devices through language-to-code transformation. Similarly, the work in~\cite{yan2025general} proposes a smart-building agent framework based on the ReAct strategy, where an LLM performs reasoning and action planning while tool calls execute actions in virtual and physical environments.

\textbf{LLMs for Sensor and Physical-World Data Interpretation}: A related line of research investigates how LLMs can interpret physical-world and sensor-derived data. The work in~\cite{xu2024penetrative} introduces the concept of ``Penetrative AI,'' where LLMs are extended beyond text-only interaction to reason about the physical world through IoT sensors and actuators, considering both textualized sensor signals and digitized numerical signals to interpret physical contexts such as user activity and physiological signals.

SensorLLM~\cite{li2024sensorllm} further investigates the alignment between motion-sensor time series and natural language. It maps sensor inputs to descriptive trend representations and introduces special tokens to preserve channel-level information for human activity recognition. These works highlight an important challenge in LLM-based sensing: numerical sensor readings must often be transformed into language-compatible representations before they can be interpreted reliably by language models.

\textbf{Edge, Cloud, and Hybrid LLM Deployment}: Deploying LLMs in IoT environments requires balancing model capability, latency, privacy, connectivity, and resources. Cloud-based LLMs provide strong reasoning and generation capabilities; however, they require continuous connectivity and may raise privacy and latency concerns. In contrast, local deployment improves data locality and responsiveness but is limited by memory and computation. Compact open-source models such as TinyLlama~\cite{zhang2024tinyllama} show that small language models can achieve useful performance with reduced resources, making them attractive for edge AI.

Several systems have been proposed to improve LLM inference under resource-constrained or distributed settings. PowerInfer-2~\cite{xue2024powerinfer} enables fast LLM inference on smartphones by using cluster scheduling. EdgeShard~\cite{zhang2024edgeshard} partitions LLMs across heterogeneous edge devices to reduce latency and improve throughput in collaborative edge computing environments. SplitLLM~\cite{mudvari2024splitllm} formulates collaborative inference between client devices and servers, using dynamic programming to allocate computation while satisfying service-level constraints. Similarly, the work in~\cite{chen2024adaptive} studies how LLM layers can be distributed between user equipment and edge nodes under wireless network variability, while HPipe~\cite{ma2024hpipe} explores pipeline parallelism for long-context LLM inference across heterogeneous commodity devices.

\textbf{Reasoning and Prompting Strategies for Local LLMs}: Prompting strategies are another important factor in LLM-based reasoning. Chain-of-thought (CoT) prompting encourages the model to generate intermediate reasoning steps before producing a final answer and has been shown to improve arithmetic, commonsense, and symbolic reasoning in large models~\cite{wei2022chain}. Such prompting is potentially useful for IoT analytics since sensor-based decisions may require comparisons against multiple limits or environmental conditions.

However, CoT can also increase output length, inference latency, and parsing difficulty, which is problematic in latency-sensitive edge settings. In structured IoT tasks, the system often requires concise, machine-readable answers rather than long explanations. These trade-offs have motivated evaluations of CoT not only in terms of reasoning accuracy, but also in terms of output reliability, latency, and deployment feasibility.

\textbf{LLM-Based Air-Quality Analytics}: Air-quality monitoring is a mature and safety-critical IoT application area. Traditional IoT-based systems integrate sensing devices, communication networks, storage, alerting, and machine-learning pipelines for prediction or anomaly detection. For example, the work in~\cite{garcia2022smart} presents an IoT-based air-quality monitoring infrastructure for industrial environments, combining compact sensing devices, LoRaWAN communication, real-time data collection, and machine-learning-based prediction.

Recently, LLMs have been applied to air-quality prediction and analysis. LLMAir~\cite{Fan2024LLMAir} adapts pre-trained LLMs to air-quality prediction by constructing spatio-temporal tokens and aligning them with the semantic space of a frozen LLM through adaptive reprogramming. Other studies use LLMs as natural-language interfaces for air-quality data access and interpretation. VayuBuddy~\cite{Patel2024VayuBuddy} presents an LLM-powered chatbot that answers stakeholder questions about air-quality data by generating and executing Python code, and evaluates multiple LLMs on curated air-quality question-answer pairs. The work in~\cite{gao2025instructor} proposes an Instructor--Worker multi-agent LLM system for wildfire-related air-quality analysis, where an Instructor agent retrieves cloud-based GIS data and delegates analysis to Worker agents for policy recommendation and health-advisory generation.

\textbf{Research Gap}: Existing studies have not sufficiently examined how prompt-side preprocessing affects the accuracy, and latency of local LLMs in comparison with cloud models. This question is important for IoT systems since local inference can improve privacy, reduce cloud dependence, and lower response latency, but compact models may struggle with raw numerical measurements~\cite{romaszewski2025through}. This paper addresses the gap by comparing raw, threshold-aware, and status-enriched prompt representations for air-quality and thermal-comfort classification across local and cloud LLMs under CoT and No-CoT inference modes.

\section{Method}\label{sec:Method}

In this work, we investigate how large language models (LLMs) can be used as an interface to a smart environment. LLMs are a natural choice for this role because they can interpret free-form user queries, integrate heterogeneous information, and generate coherent, context-aware responses. Smart environments are already deployed in many domains and are expected to become more pervasive as sensing, connectivity, and automation technologies mature, which increases the need for intuitive, language-based interaction with their data and services. In our study, we focus on air quality and thermal comfort as the physical aspect of the smart environment, inspecting it through sensors and collecting several datasets under different conditions. Using these datasets as input, we design and evaluate question answering interactions in which an LLM is asked to interpret, explain, and compare air-quality and thermal comfort measurements, thereby assessing its ability to mediate human access to smart-environment data.

We are observing the state of some environment $\mathcal{E}$, through the measurement values (data points) of $M$ measurement points across $T$ time instances:
\begin{equation}
    \mathbf{D} = \begin{bmatrix} d_{tm}\end{bmatrix}_{T\times M}\quad d_{tm}\in\mathbb{R}.
\end{equation}
Each of $M$ measurement points has a name label $l_i, i=1\dots M$.

We thus view the smart environment primarily as a source of data. In addition to raw sensor readings, we also consider derivative quantities, defined as values obtained by applying computations, algorithms, models, or external references to these readings. We argue that this possible derivative complexity can be usefully represented through two complementary aspects. First, we consider external \emph{interpretation} of measured values, for example by relating them to regulatory thresholds, health-based exposure limits, or application-specific requirements such as target ranges used in building management. Second, we consider \emph{internal processing} of the data, which could include operations such as trend analysis, anomaly detection, and event classification (e.g., identification of ventilation failures or short-term pollution episodes).

\subsection{Intepretation of data}

A single \emph{interpretation} consists of a pair of functions, indicator $f$ and thresholding $\lambda$ function. Indicator for time $t$ is defined as
\begin{equation}
    f : \mathbb{R}^m \rightarrow \mathbb{R}, \quad
    f(d_{t1}, \dots, d_{tm}) = f(\mathbf{d}_t) \in \mathbb{R}.
\end{equation}
As a concrete example, let $d = (d^\mathrm{temp}, d^\mathrm{humid}) \in \mathbb{R}^2$ be the pair of two measurement variables, denoting air temperature (in $^\circ$C) and relative humidity (in \%), respectively. We can define a thermal comfort function $f_{\text{tc}} : \mathbb{R}^2 \rightarrow \mathbb{R}$ as the NOAA simple heat-index formula:
\[
    f_{\text{tc}}(\mathbf{d}) = 1.98\,d^{\mathrm{temp}} + 0.047\,d^{\mathrm{humid}} + 24.9,
\]
where $f_{\text{tc}}$ returns the apparent temperature in $^\circ$F; higher values indicate greater heat stress.

A thresholding function $\lambda$ maps indicator values to a set of human-interpreted labels $\mathcal{L}$
\[
    \lambda : \mathbb{R} \rightarrow \mathcal{L}
\]
that maps a scalar value $x \in \mathbb{R}$ (e.g., $x = f(\mathbf{d})$) to some label $\lambda(x) \in \mathcal{L}$. As an example, consider an indoor air quality (IAQ) indicator $q \in \mathbb{R}$ based on CO$_2$ concentration (in ppm). Let $\mathcal{L} = \{\text{good}, \text{moderate}, \text{poor}, \text{very poor}\}$. A simple norm-based thresholding function can be defined as
\[
    \lambda_{\text{IAQ}}(q) =
    \begin{cases}
        \text{good},      & q \le 800,\\
        \text{moderate},  & 800 < q \le 1000,\\
        \text{poor},      & 1000 < q \le 1400,\\
        \text{very poor}, & q > 1400.
    \end{cases}
\]

\subsection{Internal processing of the data}

A single \emph{internal processing} function assigns a discrete label based on selected points of measurement history. A labelling function  produces, for each time step, a label aligned with the original time span. Let $\mathcal{Y}$ be a finite set of labels and let
\begin{equation}
    \mathbf{y} = (y_1,\dots,y_T) \in \mathcal{Y}^T
\end{equation}
denote the corresponding sequence of labels. A labelling function is any mapping
\begin{equation}
    \Lambda : \mathbb{R}^{T} \rightarrow \mathcal{Y}^T,
\end{equation}
which assigns to a given history $\mathbf{d}_m$ a label $y_t$ for each time index $t$:
\begin{equation}
    \mathbf{y} = \Lambda(\mathbf{d}_m) = (y_1,\dots,y_T).
\end{equation}

As a concrete example, consider z-score based anomaly detection for a single meter $m^\star$. Let $\mathcal{Y} = \{\text{normal}, \text{anomalous}\}$, and let $\mu_{m^\star}$ and $\sigma_{m^\star}$ denote the mean and standard deviation of the series $\{d_{t m^\star}\}_{t=1}^T$ over a reference window. For each time step $t$ we define the z-score
\begin{equation}
    z_t = \frac{d_{t m^\star} - \mu_{m^\star}}{\sigma_{m^\star}}.
\end{equation}
Given a threshold $\tau > 0$, the corresponding z-score labelling function produces
a time-aligned label sequence $\mathbf{y} = \Lambda_{\mathrm{z}}(\mathbf{D})$ with
\begin{equation}
    y_t =
    \begin{cases}
        \text{anomalous}, & \lvert z_t \rvert > \tau, \\
        \text{normal},    & \text{otherwise},
    \end{cases}
    \quad t = 1,\dots,T.
\end{equation}

\subsection{LLM queries}

Having defined numeric indicators and labelling functions, we next consider how these quantities are exposed to users through an LLM interface. In general, LLMs enable interaction with smart environments via natural-language queries that may refer to sensor values, derived indicators, or their labels, rather than requiring users to inspect raw time series or dashboards. The possible interactions span simple status checks and threshold queries, through explanations of detected events, to multi-step reasoning about control actions and hypothetical scenarios. In our setting, the LLM receives a textual description of the current air-quality and comfort indicators computed from the meter history and is asked to answer user queries about them. To reduce ambiguity and simplify evaluation, we focus on short, decision-oriented prompts that require binary answers, such as: ``Is the air quality suitable for a walk today? Answer yes or no.'', ``Are temperature and humidity comfortable indoors today? Answer yes or no.'', and ``Is it OK to go outside today? Answer yes or no.''

We have a set of $n$ questions, $\mathcal{Q}=\{q_i\}_{i=1}^{n}$, and their ground-truth answers $\mathcal{A}=\{a_i\}_{i=1}^{n}$. Let $\pi_{\mathrm{sys}}$ denote the fixed system prompt used in all experiments. Across all variants, the user-facing part of the prompt begins with a shared limits block $L_t$, which lists for each factor $l_j$ its acceptable range $[l_j^{\min}, l_j^{\max}]$ and measurement unit. For each question $q_i$, we prepare a prompt using the relevant sensor data from time $t$. The prompt is fed into an LLM, and its output is compared against the ground-truth answer $a_i$.

\begin{enumerate}
    \item \textbf{Variant A (base):} The LLM prompt is constructed as
    \begin{equation}
        P^{A}_{t,i} =
        \bigl(\pi_{\mathrm{sys}},\,L_t,\,
        l_1:d_{t1},\ldots,l_M:d_{tM},\,q_i\bigr),
    \end{equation}
    where $l_1,\ldots,l_M$ are the measurement variable names and $d_{t1},\ldots,d_{tM}$ are the measured values at time $t$.

    This variant tests the basic, native processing of numerals with language models.

    \item \textbf{Variant B:} This variant refers to numerical augmentation, and we use it to construct the LLM prompt as
    \begin{equation}
        P^{B}_{t,i} =
        \bigl(\pi_{\mathrm{sys}},\,L_t,\,\mathcal{S}(\mathbf{d}_t),\,q_i\bigr),
    \end{equation}
    where $\mathcal{S}(\mathbf{d}_t)$ denotes the textual description generated from the features of record $\mathbf{d}_t$.

    For each feature with label $l_j$ and value $d_{tj}$, we first define the limit-violation flag
    \begin{equation}
        v_{tj} =
        \begin{cases}
            1, & d_{tj}<l_j^{\min}\ \lor\ d_{tj}>l_j^{\max},\\
            0, & l_j^{\min}\leq d_{tj}\leq l_j^{\max}.
        \end{cases}
    \end{equation}

    We then construct a sentence $s_{tj}$ according to three cases:
    \begin{itemize}
        \item If $d_{tj} > l_j^{\max}$ (upper limit exceeded, $v_{tj}=1$):
        \begin{quote}
        \small
        ``$l_j$ is $d_{tj}$, exceeding the upper limit of $l_j^{\max}$ by $p_j\%$.''
        \end{quote}
        where $p_j = \lvert d_{tj} - l_j^{\max}\rvert / \lvert l_j^{\max}\rvert \times 100$.
        \item If $d_{tj} < l_j^{\min}$ (lower limit exceeded, $v_{tj}=1$):
        \begin{quote}
        \small
        ``$l_j$ is $d_{tj}$, below the lower limit of $l_j^{\min}$ by $p_j\%$.''
        \end{quote}
        where $p_j = \lvert d_{tj} - l_j^{\min}\rvert / \lvert l_j^{\min}\rvert \times 100$.
        \item If $l_j^{\min}\leq d_{tj}\leq l_j^{\max}$ (within range, $v_{tj}=0$):
        \begin{quote}
        \small
        ``$l_j$ is $d_{tj}$, within the defined range ($l_j^{\min}$--$l_j^{\max}$).''
        \end{quote}
    \end{itemize}

    Finally, the middle part of the prompt is given by the collection of all generated sentences:
    \begin{equation}
        \mathcal{S}(\mathbf{d}_t)=\{s_{t1},\ldots,s_{tM}\}.
    \end{equation}

    This variant extends the information given to the LLM with basic meta-information about signal statistics.

    \item \textbf{Variant C (improved meta-information):} The LLM prompt is constructed as in Variant B, but extended with two additional summary flags:
    \begin{equation}
        P^{C}_{t,i} =
        \bigl(\pi_{\mathrm{sys}},\,L_t,\,\mathcal{S}(\mathbf{d}_t),\,
        \mathrm{AQ}_{t},\,\mathrm{TC}_{t},\,q_i\bigr).
    \end{equation}

    The flags are computed separately for air-quality factors ($\mathcal{J}_{\mathrm{aq}}=\{j:\mathrm{pollutant}_j=1\}$) and thermal-comfort factors ($\mathcal{J}_{\mathrm{tc}}=\{j:\mathrm{pollutant}_j=0\}$) using the per-feature violation flags $v_{tj}$ from Variant B. The air-quality flag reads:
    \begin{itemize}
        \item If $\exists\,j\in\mathcal{J}_{\mathrm{aq}}:v_{tj}=1$:
        \begin{quote}
        \small
        ``Air quality safety status: unsafe (one or more pollutant limits exceeded).''
        \end{quote}
        \item Otherwise:
        \begin{quote}
        \small
        ``Air quality safety status: safe (all pollutant values within limits).''
        \end{quote}
    \end{itemize}
    The thermal-comfort flag reads:
    \begin{itemize}
        \item If $\exists\,j\in\mathcal{J}_{\mathrm{tc}}:v_{tj}=1$:
        \begin{quote}
        \small
        ``Thermal comfort status: outside recommended comfort range (discomfort likely).''
        \end{quote}
        \item Otherwise:
        \begin{quote}
        \small
        ``Thermal comfort status: within recommended comfort range.''
        \end{quote}
    \end{itemize}
\end{enumerate}

This variant extends the meta-information with a basic evaluation of the current environmental state.

\section{Dataset}\label{sec:Dataset}

\textbf{Sensor Datasets}: The evaluation uses both indoor and outdoor environmental datasets\footnote{Dataset: \url{https://doi.org/10.5281/zenodo.20783874}}. The indoor datasets correspond to three AIRWISE micro-environments: hallway, kitchen, and office. Each processed indoor dataset contains 120 sensor rows representing 20 timestamps and six parameters: temperature, humidity, IAQ proxy, and the corresponding z-score features for these variables. The IAQ proxy is derived from gas-resistance measurements and is used as an indoor air-quality indicator. The z-score features provide anomaly-oriented representations of the same environmental variables.

The outdoor datasets correspond to three city-level environments: Helsinki, Katowice, and Warsaw. Each processed outdoor dataset contains 120 sensor rows representing 20 timestamps and six parameters: carbon monoxide (CO), nitrogen dioxide ($\mathrm{NO_2}$), ozone ($\mathrm{O_3}$), fine particulate matter ($\mathrm{PM_{2.5}}$), temperature, and relative humidity. 

\textbf{Outdoor Reference Data}: The outdoor datasets are prepared from historical meteorological and air-quality data recorded in 2023. The Polish datasets are based on data from the Polish Institute of Meteorology and Water Management (IMWM)~\cite{IMGW2025} and the Chief Inspectorate of Environmental Protection, Poland~\cite{GIOS2025}. The Finnish dataset is based on data from the Finnish Meteorological Institute (FMI)~\cite{FMI2025}. The selected cities are Helsinki, Warsaw, and Katowice. For each city, the evaluated parameters are temperature (\textdegree C), relative humidity (\%), $\mathrm{NO_2}$ ($\mathrm{\mu g/m^3}$), $\mathrm{O_3}$ ($\mathrm{\mu g/m^3}$), $\mathrm{PM_{2.5}}$ ($\mathrm{\mu g/m^3}$), and CO ($\mathrm{mg/m^3}$).

Each parameter is assigned a lower and/or upper boundary according to air-quality and thermal-comfort references. For pollutants, boundary values are determined using EU air-quality standards~\cite{EU_Air_Quality_Standards}. For temperature and humidity, we consider the Heat Index (HI), based on the NOAA definition~\cite{NOAAHeatIndex2025} and the Steadman calculation scheme~\cite{NWSHeatIndex2025}, since perceived thermal stress depends on the combined effect of temperature and humidity. Conditions corresponding to a heat index above 80~\textdegree F, approximately 26.7~\textdegree C, are treated as anomalous. In addition, country-specific temperature warning levels are used: IMWM thresholds for Poland in the range of $(-15,30)$~\textdegree C~\cite{IMGWWarnings2025} and FMI thresholds for Finland in the range of $(-20,27)$~\textdegree C~\cite{FMIWarnings2025}.

\textbf{Creation of the LLM Query Dataset}: We created an LLM query dataset to evaluate whether language models can answer practical user questions from environmental sensor data. The dataset contains binary yes/no questions grouped into three task types. Type~0 questions focus on air-quality conditions, Type~1 questions focus on thermal-comfort conditions, and Type~2 questions require joint assessment of both air-quality and thermal-comfort conditions.

Example Type~0 questions include ``Is the air quality good enough to work here?'' and ``Are all air pollution measurements within safe limits?'' Type~1 questions include ``Is the indoor environment not too hot or too cold for workers?'' and ``Is the temperature and humidity comfortable for typical indoor activities?'' Type~2 questions include ``Are both air quality and thermal parameters within the limits?'' and ``Are any indoor environmental conditions outside acceptable limits, indicating that intervention is required?''

For each indoor and outdoor location, the query dataset contains 240 question-answer rows, with 80 rows for each task type. Each row includes the timestamp, question, expected answer, task type, relevant factors, and a comment describing the corresponding environmental condition. Each user query is paired with the corresponding raw sensor vector using the structured encoding format introduced in~\cite{romaszewski2025through}.

\textbf{Ground-Truth Answer Generation}: The ground-truth answer for each query is generated deterministically from the applicable threshold rules. For Type~0 questions, the answer depends on whether the air-quality-related parameters remain within their accepted limits. For Type~1 questions, the answer depends on whether temperature and humidity satisfy the thermal-comfort criteria. For Type~2 questions, the answer is positive only when both air-quality and thermal-comfort conditions are acceptable; otherwise, it indicates that at least one environmental condition requires attention.

\section{Experimental Setup}\label{sec:Experimental}

\subsection{The AIRWISE Environment}

Our prototype smart environment is AIRWISE, an Automated Indoor Regulation Wellness-Integrated Smart Environment designed for monitoring, reporting, and informing users about air-quality and thermal-comfort conditions. AIRWISE combines distributed IoT sensing, local data storage, prompt-side preprocessing, and LLM-based interpretation to support user-oriented environmental decision-making.

\begin{figure}
    \centering
    \includegraphics[width=1\linewidth]{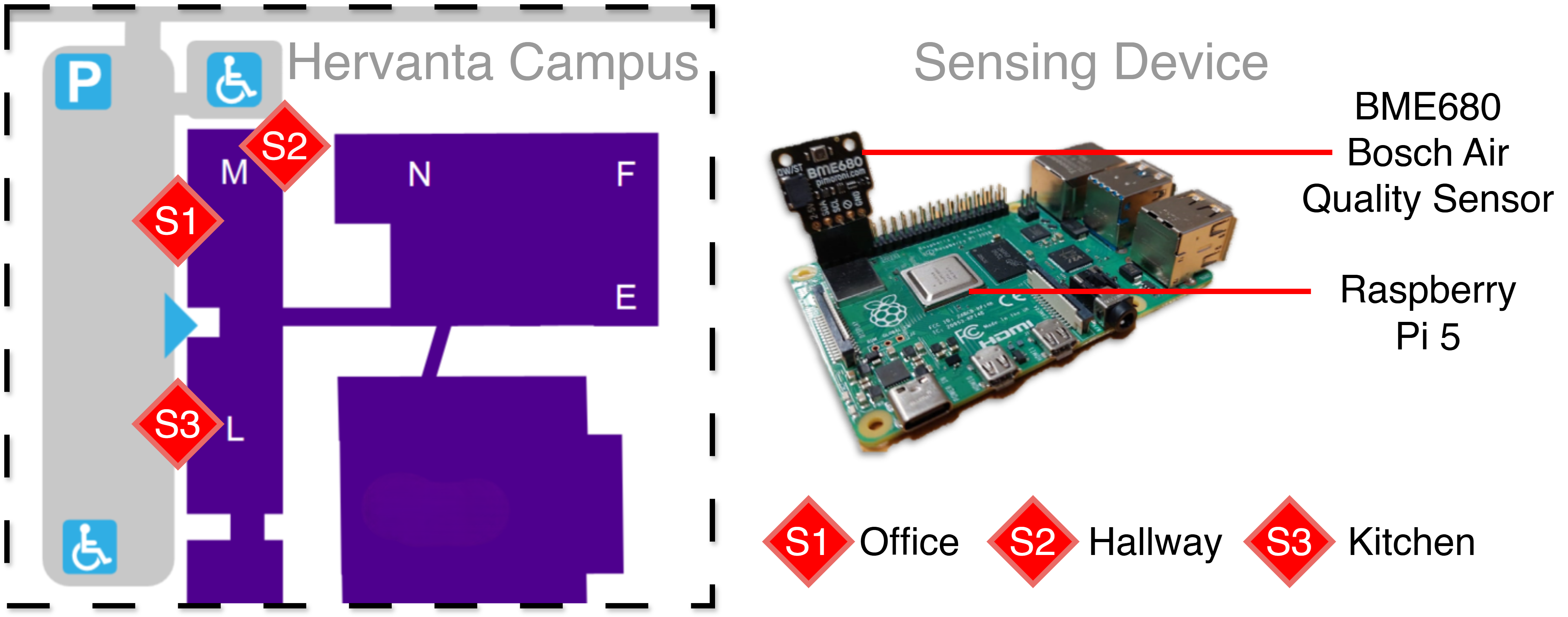}
    \caption{Sensor device and indoor deployment locations used in the AIRWISE environment.}
    \label{fig:sensor}
\end{figure}

\textbf{Sensor devices and locations:}
The indoor sensing setup consists of three Raspberry Pi nodes: one Raspberry Pi 4 and two Raspberry Pi 5 boards. Each node is equipped with a BOSCH BME680 Pimoroni air-quality sensor, which measures temperature, pressure, relative humidity, and gas resistance once per second~\cite{bosch2024bme680}. The readings are buffered locally on the Raspberry Pi devices. The Raspberry Pi 4 provides 64~GB of onboard storage, while each Raspberry Pi 5 provides 256~GB of storage and 16~GB of RAM. The sensor nodes are deployed in three indoor micro-environments at Tampere University, Hervanta Campus: kitchen, hallway, and office. Fig.~\ref{fig:sensor} shows the sensor device and deployment locations.

\subsection{Reference Limits and Input Prompts}

The prompt construction process uses environmental reference limits to convert raw sensor readings into threshold-aware textual representations. The limit values are derived from air-quality and comfort references, including EU air-quality standards~\cite{EU_Air_Quality_Standards} and ASHRAE ventilation and indoor air-quality guidance~\cite{ASHRAE_62_1_2022}. 

Each query is paired with the corresponding sensor record and transformed according to the three prompt variants (Var.~A--Var.~C) defined in Section~\ref{sec:Method}. This setup enables a controlled evaluation of how progressively enriched prompt representations affect model performance.

\subsection{LLM Inference Modes}

We evaluate two inference modes. In the No-CoT mode, the system prompt instructs the model to return exactly one lowercase binary answer: \texttt{yes} or \texttt{no}. In the CoT mode, the system prompt requests a structured XML response composed of a \texttt{<scratchpad>} and an \texttt{<action>} block; the final binary decision is extracted from the \texttt{<label>} element inside \texttt{<action>}. The scratchpad enforces a three-step reasoning chain: the model first identifies the question scope (air quality, thermal comfort, or overall conditions), then enumerates only the limit violations relevant to that scope, and finally justifies its yes/no decision based on those findings alone---out-of-scope factors are explicitly ignored. This structured decomposition is intended to reduce spurious answers caused by irrelevant sensor readings while providing a human-readable reasoning trace alongside each decision.

For each model, dataset, prompt variant, and inference mode, the system sends a prompt consisting of a fixed system instruction and a user message containing the environmental measurements, reference-limit information, and user query. The temperature is fixed at 0.2 in all experiments to reduce response variability. For local inference, a warm-up query is executed before batch evaluation to avoid including first-load model latency in the measured experiment rows.

For every evaluated row, we store the raw model response, parsed binary prediction, expected answer, correctness flag, task type, and response latency. Latency is measured as the elapsed time between sending the prompt and receiving the model output.

\subsection{Model Selection and Hardware}

We evaluate five locally deployed models: Granite 4.1 3B, Llama 3.2 3B Instruct, Llama 3.1 8B, Ministral 3 3B, and Phi-4 Mini. 
These models were selected since they represent state-of-the-art performers within their parameter classes, span distinct model families, 
and cover a range of parameter scales suited for resource-constrained deployment. 
Local inference is performed using Ollama\footnote{\url{https://ollama.com/}}, where each request contains a system message, a user message, and the fixed temperature setting. We have conducted all our local model experiments on a common laptop with an RTX 2060 6GB GPU, Intel Core i7 10750H, and 32 GB DDR4.

For cloud-based comparison, we evaluate five cloud models using API-based inference, namely gemini-2.5-flash, gemini-3.1-flash-lite, gpt-5.4, gpt-5.4-mini, and gpt-5.4-nano. 
The cloud models are treated as high-capacity reference systems for evaluating the effect of prompt enrichment and inference mode against locally deployed models. All local and cloud models are evaluated using the same indoor and outdoor query datasets, prompt variants, and CoT/No-CoT settings.

\subsection{Evaluation Metrics}

The primary evaluation metrics are classification accuracy, parse rate, and mean latency. Accuracy measures the fraction of correct predictions, parse rate measures the fraction of model responses that follow the required output format, and mean latency measures the average response time per evaluated row. In addition, balanced accuracy, F1 score, per-task-type accuracy, and total latency are computed.

\section{Results}\label{sec:Results}

The evaluation covers 180 result files amounting to approximately 86\,400 individual inference rows, spanning five local models (\textit{granite4.1-3b}, \textit{Llama-3.2-3B}, \textit{llama3.1-8b}, \textit{ministral-3-3b}, and \textit{phi4-mini}) and five cloud models (\textit{gemini-2.5-flash}, \textit{gemini-3.1-flash-lite}, \textit{gpt-5.4}, \textit{gpt-5.4-mini}, and \textit{gpt-5.4-nano}), evaluated on three prompt variants and two inference modes (No-CoT and CoT). Throughout this section, Variant~A refers to raw sensor values, Variant~B to threshold-aware descriptions, and Variant~C to environmental summary flags, respectively, as defined in Section~\ref{sec:Method}. Three axes are reported: classification accuracy (together with a balanced-accuracy variant), response parse rate, and per-row latency, across three indoor datasets (hallway, kitchen, office) and three outdoor datasets (Helsinki, Katowice, Warsaw).

\vspace{-11pt}

\begin{figure*}[!t]
  \centering
  \includegraphics[width=0.85\textwidth]{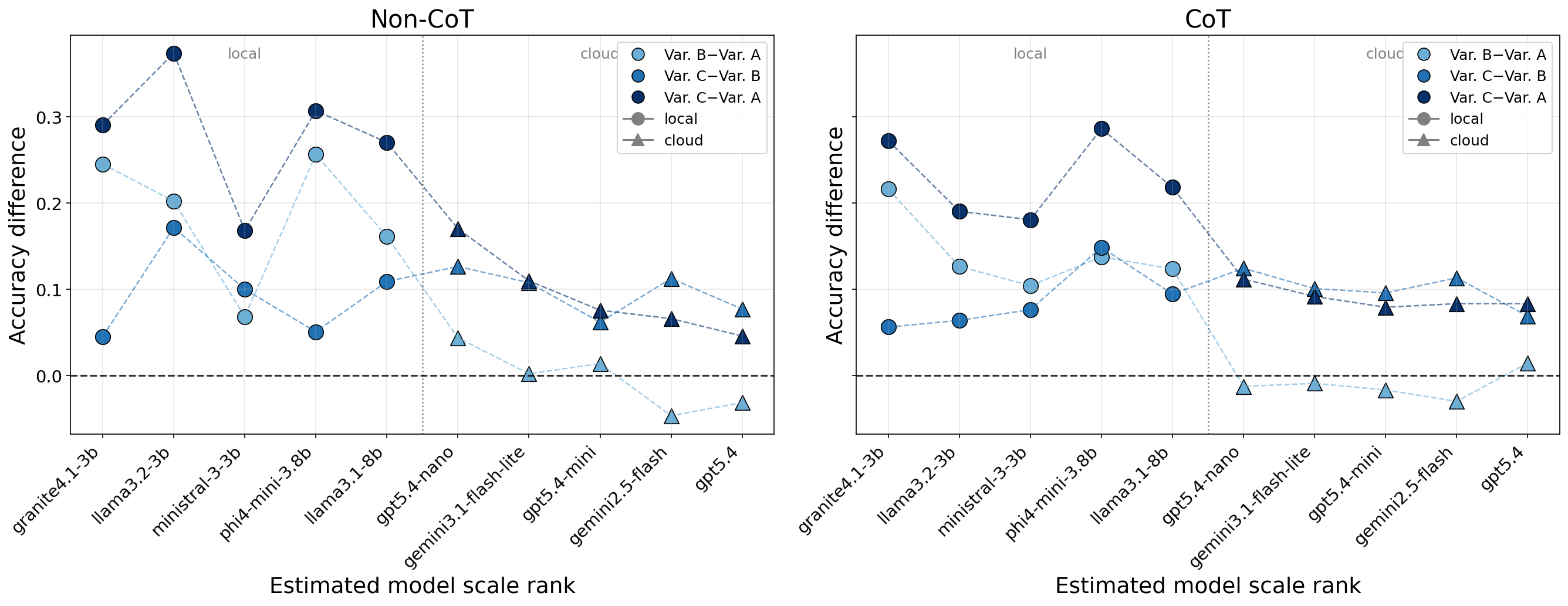}
  \caption{Accuracy differences between experiment variants plotted against estimated model scale rank. Circles: local models; triangles: cloud models. Left: Non-CoT; right: CoT. The dashed horizontal line shows the threshold above which a desired positive difference occurs. }
  \label{fig:acc_diff}
\end{figure*}

\begin{figure*}[!t]
  \centering
  \includegraphics[width=0.85\textwidth]{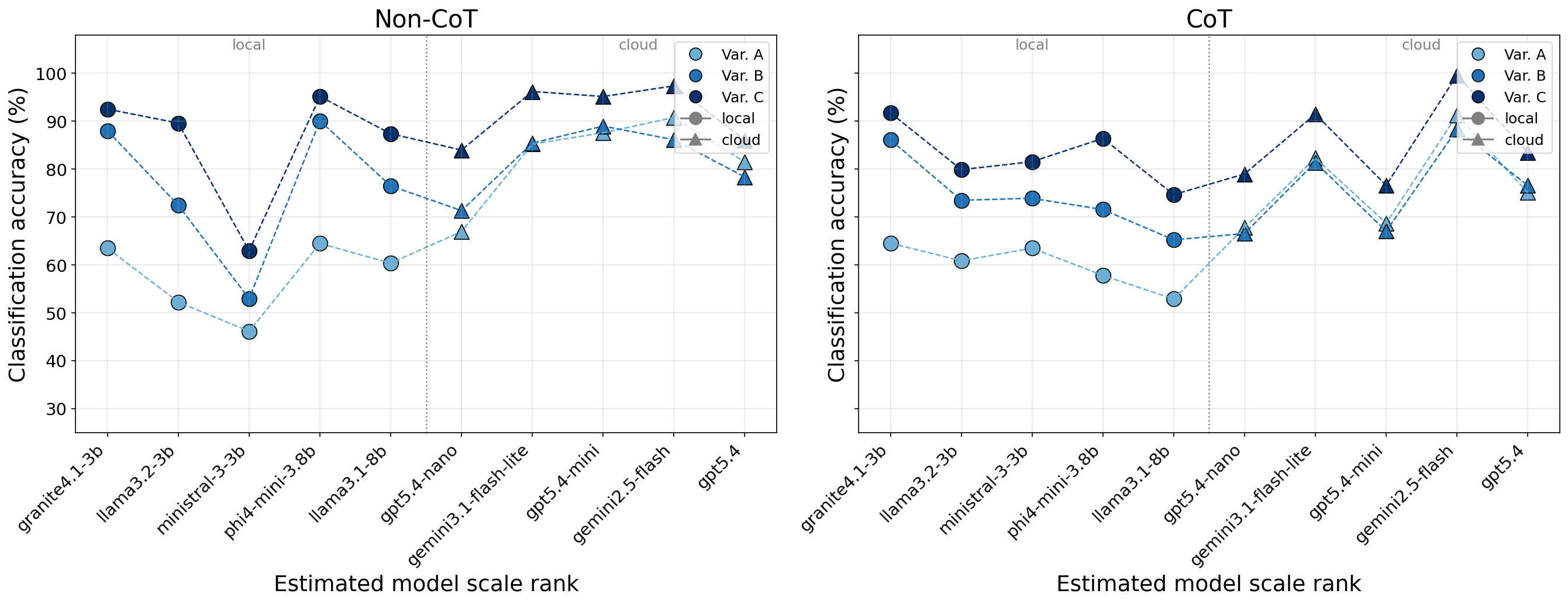}
  \caption{Absolute classification accuracy per model and experiment variant. Local models (circles) and cloud models (triangles) are separated by a vertical dotted line. In the Non-CoT panel, local and cloud accuracy values converge by Var.~C (except for \textit{ministral-3-3b}), confirming that prompt enrichment eliminates most of the accuracy gap between edge and cloud deployment.}
  \label{fig:acc_abs}
\end{figure*}

\subsection{Aggregate Analysis}\label{sec:agg}

We report three complementary aggregate views that reveal systematic trends. \textbf{Experiment-variant} aggregation averages Var.~A--Var.~C over all models and all locations. \textbf{Model aggregation} averages each model over all experiments and locations, separating CoT and No-CoT. \textbf{Latency aggregation} summarizes mean per-row inference time per deployment setting and inference mode. Detailed per-model and per-location results are provided in Section~\ref{subsec:detailed_results}.

\begin{tcolorbox}[%
  colback=blue!6!white, colframe=blue!35!white, arc=2mm, boxrule=0.6pt,
  title={\textbf{Key Findings}}, fonttitle=\bfseries,
  before upper={\parskip4pt}]

\textbf{Finding 1 --- Prompt enrichment is the dominant accuracy driver.}
Averaged over all five local models and all indoor locations, No-CoT accuracy increases by $+30.8$\,pp from Var.~A (50.9\%) to Var.~C (81.7\%); the outdoor gain is $+25.6$\,pp. Threshold-aware context (Variant~B) accounts for the largest step (+20.6\,pp indoor; +16.7\,pp outdoor); the explicit summary flag of Variant~C adds a further, smaller gain (see Tab.~\ref{tab:agg_exp}). Fig.~\ref{fig:acc_diff} shows that local models (circles) systematically obtain larger Var.~C--Var.~A gains than cloud models (triangles) in No-CoT mode, with smaller local models benefiting proportionally more.

\medskip
\textbf{Finding 2 --- Under optimal prompting, local models reach cloud-level accuracy.}
Under Var.~C No-CoT, \textit{phi4-mini} achieves 97.4\% (indoor) and \textit{llama3.1-8b} achieves 95.0--97.9\% across outdoor cities, placing them within 0--2 percentage points of the best cloud models (\textit{gemini-2.5-flash} at 99.2\% indoor; \textit{gemini-3.1-flash-lite} at 97.5--98.8\% outdoor). Fig.~\ref{fig:acc_abs} shows this convergence: the accuracy gradient across model scales collapses from $\sim$40\,pp in Var.~A to $<$10\,pp in Var.~C for No-CoT inference.
\end{tcolorbox}

\textbf{Experiment-variant aggregation.}
Table~\ref{tab:agg_exp} shows large, monotonic accuracy gains for local models. Indoors, the local No-CoT mean rises from 50.9\% (Var.~A) to 71.5\% (Var.~B) and 81.7\% (Var.~C), a cumulative gain of $+30.8$\,pp; outdoors the gain is $+25.6$\,pp (63.7\%$\to$89.3\%). Var.~B carries the largest step in both environments, confirming that threshold-aware context is the most impactful prompt component. Cloud models show a different pattern: indoor No-CoT is nearly unchanged between Var.~A (74.5\%) and Var.~B (74.3\%), then rises sharply in Var.~C (88.6\%), while outdoor cloud accuracy starts from a high baseline ($\geq$90\%) and improves further but more modestly. The CoT direction is consistent: after prompt enrichment, No-CoT outperforms CoT in all four deployment settings (Var.~B and Var.~C). The cumulative gain from Var.~A to Var.~C is lower for cloud models than for local ones in all deployment settings.  

\begin{table}[!t]
\centering
\scriptsize
\setlength{\tabcolsep}{3.2pt}
\renewcommand{\arraystretch}{1.08}
\caption{%
Experiment-variant aggregation. Mean classification accuracy (\%).
$\Delta_{A\to C}$: gain from Var.~A to Var.~C in
percentage points (pp). \textbf{N}: No-CoT; \textbf{C}: CoT.}
\label{tab:agg_exp}
\begin{tabular*}{\columnwidth}{@{\extracolsep{\fill}} L{0.20\columnwidth} c r r r r @{}}
\toprule
\textbf{Setting} & \textbf{Mode}
  & \textbf{Var.~A} & \textbf{Var.~B} & \textbf{Var.~C}
  & \boldmath{$\Delta_{A\to C}$} \\
\midrule
\multirow{2}{*}{\makecell[l]{Local\\Indoor}}
  & N & 50.9 & 71.5 & 81.7 & $+30.8$ \\
  & C & 57.0 & 67.9 & 78.8 & $+21.9$ \\
\midrule
\multirow{2}{*}{\makecell[l]{Local\\Outdoor}}
  & N & 63.7 & 80.4 & 89.3 & $+25.6$ \\
  & C & 62.8 & 80.2 & 86.8 & $+24.0$ \\
\midrule
\multirow{2}{*}{\makecell[l]{Cloud\\Indoor}}
  & N & 74.5 & 74.3 & 88.6 & $+14.2$ \\
  & C & 62.0 & 62.9 & 76.6 & $+14.6$ \\
\midrule
\multirow{2}{*}{\makecell[l]{Cloud\\Outdoor}}
  & N & 90.3 & 89.7 & 94.8 & $+4.5$ \\
  & C & 92.0 & 88.9 & 95.4 & $+3.4$ \\
\bottomrule
\end{tabular*}
\renewcommand{\arraystretch}{1}
\end{table}

\textbf{Model aggregation.}
Table~\ref{tab:agg_model} reports per-model mean accuracy (averaged over all three experiments and all locations) together with $\Delta$\,CoT (CoT$-$No-CoT in pp), which quantifies the model-specific CoT benefit or penalty. Among local models, only \textit{ministral-3-3b} shows a large positive $\Delta$\,CoT ($+26.2$\,pp) for indoor datasets, concentrated in Var.~A where CoT compensates for missing threshold context; all other local models show neutral or negative deltas. All five cloud models show negative $\Delta$\,CoT, with \textit{gpt-5.4-mini} recording the largest penalty ($-37.8$\,pp). For outdoor datasets, CoT is sometimes beneficial for cloud (\textit{gemini-2.5-flash}: $+5.4$\,pp; \textit{gemini-3.1-flash-lite}: $+0.7$\,pp) and local models (\textit{Llama-3.2-3B}: $+8.9$\,pp; \textit{ministral-3-3b}: $+11.8$\,pp). Overall, No-CoT is the better default in 13 of 20 model-environment combinations.

\begin{table}[!t]
\centering
\scriptsize
\setlength{\tabcolsep}{3.0pt}
\renewcommand{\arraystretch}{1.08}
\caption{%
Per-model mean classification accuracy (\%).
$\Delta$\,C: Difference of CoT $-$ No-CoT in pp. L: local; Cl: cloud.
\textbf{N}: No-CoT; \textbf{C}: CoT.}
\label{tab:agg_model}
\begin{tabular*}{\columnwidth}{@{\extracolsep{\fill}} L{0.35\columnwidth} c r r r r r r @{}}
\toprule
 & & \multicolumn{3}{c}{\textbf{Indoor}} & \multicolumn{3}{c}{\textbf{Outdoor}} \\
\cmidrule(lr){3-5}\cmidrule(lr){6-8}
\textbf{Model} & \textbf{T}
  & \textbf{N} & \textbf{C} & \boldmath{$\Delta$\,C}
  & \textbf{N} & \textbf{C} & \boldmath{$\Delta$\,C} \\
\midrule
granite4.1-3b        & L & 75.1 & 76.0 & $+0.8$  & 87.5 & 85.6 & $-1.9$  \\
Llama-3.2-3B         & L & 70.5 & 61.6 & $-8.9$  & 72.4 & 81.2 & $+8.9$  \\
llama3.1-8b          & L & 63.4 & 53.4 & $-10.0$ & 86.1 & 75.0 & $-11.0$ \\
ministral-3-3b       & L & 52.7 & 78.9 & $+26.2$ & 55.3 & 67.0 & $+11.8$ \\
phi4-mini            & L & 78.6 & 69.6 & $-8.9$  & 87.9 & 74.2 & $-13.7$ \\
\midrule
gemini-2.5-flash      & Cl & 93.7 & 91.5 & $-2.2$  & 89.2 & 94.6 & $+5.4$ \\
gemini-3.1-flash-lite & Cl & 81.5 & 73.1 & $-8.4$  & 96.4 & 97.0 & $+0.7$ \\
gpt-5.4               & Cl & 68.8 & 64.4 & $-4.4$  & 95.1 & 92.3 & $-2.8$ \\
gpt-5.4-mini          & Cl & 90.4 & 52.6 & $-37.8$ & 90.7 & 88.9 & $-1.8$ \\
gpt-5.4-nano          & Cl & 61.3 & 54.5 & $-6.8$  & 86.8 & 87.6 & $+0.8$ \\
\bottomrule
\end{tabular*}
\renewcommand{\arraystretch}{1}
\end{table}

\textbf{Latency aggregation.}
Table~\ref{tab:agg_latency} reveals a clear four-level hierarchy. Local No-CoT is the fastest configuration at $\overline{L} = 0.216$--$0.217$\,s per row; cloud No-CoT is intermediate at 1.41--1.48\,s; cloud CoT rises to 2.41--2.72\,s; and local CoT is the slowest at 3.59--3.81\,s. Local CoT is therefore 16--18$\times$ slower than local No-CoT (in terms of aggregate latency ratio), with no aggregated accuracy benefit under enriched-prompt conditions (compare with Tab.~\ref{tab:agg_exp}). Prompt enrichment itself adds negligible latency: comparing Var.~A to Var.~C within the local No-CoT row shows an increase of only $\Delta_{A\to C} = +0.017$\,s (indoor) and $+0.027$\,s (outdoor), confirming that inference time scales with output tokens, not input length.

\begin{table}[!t]
\centering
\scriptsize
\setlength{\tabcolsep}{3.0pt}
\renewcommand{\arraystretch}{1.08}
\caption{%
Latency aggregation. Mean per-row latency (seconds).
$\Delta_{A\to C}$: change from Var.~A to Var.~C (s).
$\overline{L}$: mean over all three experiments.
\textbf{N}: No-CoT; \textbf{C}: CoT.}
\label{tab:agg_latency}
\begin{tabular*}{\columnwidth}{@{\extracolsep{\fill}} L{0.22\columnwidth} c r r r r r @{}}
\toprule
\textbf{Setting} & \textbf{Mode}
  & \textbf{Var.~A} & \textbf{Var.~B} & \textbf{Var.~C}
  & \boldmath{$\Delta_{A\to C}$}
  & \boldmath{$\overline{L}$} \\
\midrule
\multirow{2}{*}{\makecell[l]{Local\\Indoor}}
  & N & 0.209 & 0.214 & 0.226 & $+0.017$ & 0.216 \\
  & C & 3.735 & 3.438 & 3.599 & $-0.136$ & 3.591 \\
\midrule
\multirow{2}{*}{\makecell[l]{Local\\Outdoor}}
  & N & 0.202 & 0.221 & 0.229 & $+0.027$ & 0.217 \\
  & C & 4.080 & 3.704 & 3.659 & $-0.421$ & 3.814 \\
\midrule
\multirow{2}{*}{\makecell[l]{Cloud\\Indoor}}
  & N & 1.293 & 1.466 & 1.459 & $+0.166$ & 1.406 \\
  & C & 2.862 & 2.617 & 2.691 & $-0.171$ & 2.723 \\
\midrule
\multirow{2}{*}{\makecell[l]{Cloud\\Outdoor}}
  & N & 1.492 & 1.523 & 1.417 & $-0.075$ & 1.478 \\
  & C & 2.504 & 2.337 & 2.374 & $-0.130$ & 2.405 \\
\bottomrule
\end{tabular*}
\renewcommand{\arraystretch}{1}
\end{table}

\subsection{Effect of Prompt Enrichment and Inference Mode}\label{sec:enrichment}

Table~\ref{tab:optimal} gives the best-performing model and configuration for each deployment setting. The remainder of this subsection explains the mechanisms underlying these optimal settings.

\begin{table}[!t]
\centering
\scriptsize
\setlength{\tabcolsep}{4pt}
\renewcommand{\arraystretch}{1.1}
\caption{%
Optimal model and configuration per deployment setting at Var.~C No-CoT.
Accuracy is averaged over all locations.}
\label{tab:optimal}
\begin{tabular*}{\columnwidth}{@{\extracolsep{\fill}} L{0.24\columnwidth} L{0.34\columnwidth} c c r @{}}
\toprule
\textbf{Setting} & \textbf{Model} & \textbf{Mode} & \textbf{Var.} & \textbf{Acc.\,\%} \\
\midrule
Local indoor  & phi4-mini             & No-CoT & C & 97.4 \\
Local outdoor & llama3.1-8b           & No-CoT & C & 96.8 \\
Cloud indoor  & gemini-2.5-flash      & No-CoT & C & 99.2 \\
Cloud outdoor & gemini-3.1-flash-lite & No-CoT & C & 98.1 \\
\bottomrule
\end{tabular*}
\renewcommand{\arraystretch}{1}
\end{table}

Prompt enrichment consistently improves performance across deployment settings. For local models, the largest gain is obtained when moving from Var.~A to Var.~B, increasing mean accuracy by +20.6\,pp for indoor datasets and +16.7\,pp for outdoor ones under No-CoT mode. Adding the summary state flag in Var.~C yields a further improvement of +10.2\,pp indoors and +8.9\,pp outdoors. Moving from Var.~A to Var.~C increases input prompt length by only $\approx$34\,\% (from 326 to 437 tokens on average) but leaves output token count at $\approx$1~token (\textit{yes}/\textit{no}), so latency and energy cost are essentially unchanged (see Section~\ref{sec:cost}).

CoT is conditionally useful in the low-context Var.~A setting, where explicit reasoning partially compensates for missing threshold information: the aggregate local indoor mean rises from 50.9\% (No-CoT) to 57.0\% (CoT) in Var.~A (see Tab.~\ref{tab:agg_exp}). Once enriched prompts are provided, No-CoT matches or exceeds CoT for nearly all models (except for \textit{ministral-3-3b} and Gemini models, see Fig.~\ref{fig:acc_abs}).

 The Pareto-optimal configuration for local deployment is therefore Var.~C No-CoT: it delivers near-cloud accuracy at sub-0.3\,s latency with zero parse failures and negligible compute overhead relative to Var.~A. For cloud deployment, Var.~C No-CoT is equally optimal; CoT provides marginal outdoor accuracy gains for Gemini models only and should not be used as a default.

\subsection{Error Analysis}\label{sec:errors}

Table~\ref{tab:error_compact} reports unparsed-response and wrong-prediction rates for all ten models in CoT mode. A critical baseline: No-CoT produces \emph{zero} unparsed responses across all 47\,520 No-CoT inference rows; the parse problem is strictly confined to chain-of-thought mode. Three failure mechanisms account for all observed error types.

\begin{table}[!t]
\centering
\scriptsize
\setlength{\tabcolsep}{3.0pt}
\renewcommand{\arraystretch}{1.08}
\caption{%
CoT mode error rates.
\textit{Unp.\%}: unparsed rate; \\ \textit{Wr.\%}: wrong-prediction rate
(unparsed counted as wrong). \\
Cl: cloud; Lo: local.}
\label{tab:error_compact}
\begin{tabular*}{\columnwidth}{@{\extracolsep{\fill}} L{0.34\columnwidth} c r r r r @{}}
\toprule
 & & \multicolumn{2}{c}{\textbf{Indoor CoT}} & \multicolumn{2}{c}{\textbf{Outdoor CoT}} \\
\cmidrule(lr){3-4}\cmidrule(lr){5-6}
\textbf{Model} & \textbf{T}
  & \textbf{Unp.\%} & \textbf{Wr.\%}
  & \textbf{Unp.\%} & \textbf{Wr.\%} \\
\midrule
gpt-5.4-nano         & Cl &  9.1 & 45.5 &  8.5 & 12.4 \\
gpt-5.4-mini         & Cl &  0.0 & 47.4 &  3.9 & 11.1 \\
gpt-5.4              & Cl &  0.4 & 35.6 & 10.5 &  7.7 \\
gemini-3.1-flash-lite & Cl &  0.0 & 26.9 &  3.3 &  3.0 \\
gemini-2.5-flash     & Cl &  0.0 &  8.5 &  0.6 &  5.4 \\
\midrule
granite4.1-3b        & Lo &  1.9 & 24.0 &  6.7 & 14.4 \\
Llama-3.2-3B         & Lo &  5.9 & 38.4 &  5.2 & 18.8 \\
llama3.1-8b          & Lo &  3.3 & 46.6 &  2.9 & 25.0 \\
ministral-3-3b       & Lo & 11.5 & 21.1 &  5.1 & 33.0 \\
phi4-mini            & Lo & 12.8 & 30.4 & 13.6 & 25.8 \\
\bottomrule
\end{tabular*}
\renewcommand{\arraystretch}{1}
\end{table}

\textbf{Answer--reasoning inversion (polarity-related bias).}
The dominant CoT failure across both cloud and local models is answer--reasoning inversion: during CoT-enforced reasoning, the model correctly identifies limit violations and their magnitudes, but the final label contradicts this diagnosis. A plausible explanation is a polarity-related bias: the phrase ``conditions are not within limits'' is strongly associated with a negative label in safety-oriented text corpora, so when the question asks the factual direction (``is $X$ \emph{above} limits?'') the model produces the correct evidence but the wrong answer. This pattern accounts for 7 of 10 sampled wrong responses from \textit{gpt-5.4-nano} (indoor CoT: 45.5\% wrong), is independently observed in \textit{gemini-3.1-flash-lite} (indoor CoT: 26.9\% wrong vs.\ 18.5\% No-CoT), and appears in a selective-evidence variant in \textit{phi4-mini}: the model correctly identifies a limit violation in the scratchpad, then discounts it with a hedge about occupancy levels and emits the wrong label. In all observed cases this is a semantic polarity failure, not a comprehension failure --- the model understands the sensor data correctly.

\textbf{Scope and arithmetic errors.}
Two subtypes are observed. \emph{Scope over-expansion}: \textit{gpt-5.4} includes thermal comfort parameters (humidity, temperature) when answering air-pollution questions, because its scratchpad considers all environmental factors regardless of question domain. \textit{Llama-3.2-3B} exhibits an analogous indoor variant, citing temperature z-scores as grounds for failing an air-pollution classification. \emph{Threshold direction inversion}: \textit{ministral-3-3b} reports ``O$_3$: 63.45\,$\mu$g\,m$^{-3}$ exceeds limit of 120.0\,$\mu$g\,m$^{-3}$''~--- a factual arithmetic error in which the threshold value is read correctly but the comparison direction is inverted, fabricating a limit violation that does not exist. This is distinct from the polarity failures above: here the model does not understand the data correctly.

\textbf{XML parse failures.}
Inspection of malformed outputs indicates that the dominant cause of parse failures is the inclusion of unescaped comparison operators (\texttt{>} and \texttt{<}) inside \texttt{<scratchpad>} elements: human-readable comparisons such as \texttt{88.00\% > 80.0\%} are syntactically invalid XML, causing parsing to fail. A secondary cause is response truncation in models with verbose scratchpads, which exhaust the output token budget before the \texttt{<label>} element. Counterintuitively, the highest cloud unparsed rates are observed for \textit{gpt-5.4} (outdoor CoT: 10.5\%) and \textit{gpt-5.4-nano} (8.5--9.1\%), while both Gemini models remain at about 0\% for indoor datasets and below 4\% for outdoor ones. Among local models, \textit{phi4-mini} (12.8--13.6\%) and \textit{ministral-3-3b} (11.5\%) are the worst, whereas \textit{llama3.1-8b} achieves only 2.9--3.3\% despite being the largest local model. These results suggest that format compliance and reasoning correctness are largely independent capabilities.

\subsection{Computational Cost and Energy Efficiency}\label{sec:cost}

Token counts were computed using the \textit{cl100k\_base} tokeniser from the OpenAI tiktoken library and applied to every result CSV \cite{openai_tiktoken}. Prior work suggests that tokeniser choice has only a limited impact on English-language tasks, with different modern subword tokenisers yielding largely comparable downstream performance on English text \cite{lotz2025beyond}. Cloud API costs use list prices accessed on 2026-06-18~\cite{openai_api_pricing,gemini_pricing} (gpt-5.4: \$2.50\,/\,\$15.00 per million input\,/\,output tokens; gpt-5.4-mini: \$0.75\,/\,\$4.50; gpt-5.4-nano: \$0.20\,/\,\$1.25; gemini-2.5-flash: \$0.30\,/\,\$2.50; gemini-3.1-flash-lite: \$0.25\,/\,\$1.5). Local electricity cost is estimated as $E = t_\mathrm{infer} \times P_\mathrm{GPU}$ at $P_\mathrm{GPU} = 80$\,W (according to the GPU engine specs of NVIDIA GeForce RTX 2060 used in this study, \url{https://laptopmedia.com/video-card/nvidia-geforce-rtx-2060-laptop-80w/}, accessed on 2026-06-18 ) and \$0.13\,kWh$^{-1}$ (the average residential electricity price in Finland in September 2025 \cite{globalpetrolprices_finland_2025}).

\textbf{Prompt enrichment is free in compute terms.}
Table~\ref{tab:phi4_configs} shows that moving from Var.~A to Var.~C for \textit{phi4-mini} increases input tokens by $\approx$34\,\% (321$\to$422 indoor; 331$\to$453 outdoor) while output tokens remain at $\approx$1 and average latency rises by about 0.01\,s per row. The accuracy gain is $+47.4$\,pp (indoor, 50.0\%$\to$97.4\%). Enabling CoT on Var.~A expands output tokens on average by $156\times$ (to 141--172) and raises average latency approx. 17$\times$ for indoor datasets and 23$\times$ for outdoor ones, increasing GPU energy per 1\,000 calls on average from $\approx$4.6\,Wh to $\approx$90\,Wh, yet accuracy under Var.~A CoT (56.5\% indoor, 59.0\% outdoor) remains well below Var.~C No-CoT in every environment. 

\begin{table}[!t]
\centering
\scriptsize
\setlength{\tabcolsep}{3.0pt}
\renewcommand{\arraystretch}{1.08}
\caption{%
\textit{phi4-mini} across three configurations.
In: indoor; Out: outdoor.
Energy per 1\,000 calls at 80\,W; electricity at \$0.13\,kWh$^{-1}$.}
\label{tab:phi4_configs}
\begin{tabular*}{\columnwidth}{@{\extracolsep{\fill}} L{0.07\columnwidth} L{0.09\columnwidth} L{0.11\columnwidth} r r r r r r @{}}
\toprule
\textbf{Env.} & \textbf{Var.} & \textbf{Mode}
  & \textbf{In} & \textbf{Out} & \textbf{Lat.}
  & \textbf{Acc} & \textbf{Wh} & \textbf{\$/1K} \\
  & & & \textbf{tok} & \textbf{tok} & \textbf{(s)}
  & \textbf{(\%)} & \textbf{/1K} & \textbf{(elec.)} \\
\midrule
In  & A & No-CoT & 321 &   1 & 0.21 & 50.0 &  4.75 & \num{6.17e-4} \\
In  & C & No-CoT & 422 &   1 & 0.22 & 97.4 &  4.94 & \num{6.43e-4} \\
In  & A & CoT    & 427 & 141 & 3.56 & 56.5 & 79.02 & 0.0103 \\
Out & A & No-CoT & 331 &   1 & 0.20 & 78.9 &  4.42 & \num{5.75e-4} \\
Out & C & No-CoT & 453 &   1 & 0.23 & 92.9 &  5.04 & \num{6.56e-4} \\
Out & A & CoT    & 433 & 172 & 4.56 & 59.0 & 101.4 & 0.0132 \\
\bottomrule
\end{tabular*}
\renewcommand{\arraystretch}{1}
\end{table}

\textbf{Local vs.\ cloud cost differential.}
Table~\ref{tab:summary_cost} places \textit{phi4-mini} electricity cost against all cloud API prices at Var.~C No-CoT. The local solution costs approximately \$\num{6.43e-4}\,/\,1\,000 calls (indoor) --- 134$\times$ lower than the most affordable cloud option (\textit{gpt-5.4-nano} at \$0.086) and 499$\times$ lower than \textit{gpt-5.4-mini} (\$0.321). At 100\,000 daily sensor readings, the API bill for \textit{gpt-5.4-mini} reaches \$32.1\,/\,day versus \$0.064\,/\,day in electricity for \textit{phi4-mini}. The accuracy gap between \textit{phi4-mini} and cloud models at Var.~C No-CoT is at most 1.8\,pp for indoor datasets (97.4\% vs.\ 99.2\% for \textit{gemini-2.5-flash}) and 5.2\,pp outdoors (92.9\% vs.\ 98.1\% for \textit{gemini-3.1-flash-lite}), while a cost of \textit{phi4-mini} is 166--210$\times$ lower than the cost of those best-performing Gemini models. 

\begin{table}[!t]
\centering
\scriptsize
\setlength{\tabcolsep}{3.0pt}
\renewcommand{\arraystretch}{1.08}
\caption{%
Accuracy and cost at Var.~C No-CoT. Local \textit{phi4-mini} cost is
electricity only (\$0.13\,kWh$^{-1}$, 80\,W GPU) \\ Cloud cost uses list API
prices as of 2026-06-18~\cite{openai_api_pricing,gemini_pricing}.}
\label{tab:summary_cost}
\begin{tabular*}{\columnwidth}{@{\extracolsep{\fill}} L{0.34\columnwidth} l r r r @{}}
\toprule
\textbf{Model} & \textbf{Env.}
  & \textbf{Acc.\,\%} & \textbf{\$/1K calls} & \textbf{Wh/1K} \\
\midrule
phi4-mini (local)    & In  & 97.4 & \num{6.43e-4} & 4.94 \\
gemini-3.1-flash-lite & In  & 94.3 & 0.107 & --- \\
gpt-5.4-nano         & In  & 79.2 & 0.086 & --- \\
gemini-2.5-flash     & In  & 99.2 & 0.129 & --- \\
gpt-5.4-mini         & In  & 95.8 & 0.321 & --- \\
gpt-5.4              & In  & 74.7 & 1.070 & --- \\
\midrule
phi4-mini (local)    & Out & 92.9 & \num{6.56e-4} & 5.04 \\
gemini-3.1-flash-lite & Out & 98.1 & 0.115 & --- \\
gpt-5.4-nano         & Out & 88.8 & 0.092 & --- \\
gemini-2.5-flash     & Out & 95.6 & 0.138 & --- \\
gpt-5.4-mini         & Out & 94.4 & 0.344 & --- \\
gpt-5.4              & Out & 97.4 & 1.147 & --- \\
\bottomrule
\end{tabular*}
\renewcommand{\arraystretch}{1}
\end{table}

\subsection{Detailed Per-Model and Per-Location Results}
\label{subsec:detailed_results}

This subsection provides the complete per-model and per-location accuracy and latency results for all ten evaluated models across the three prompt variants (Var.~A--Var.~C), two inference modes (No-CoT and CoT), three indoor datasets (hallway, kitchen, office), and three outdoor datasets (Helsinki, Katowice, Warsaw). These detailed tables complement the aggregate accuracy, latency, error, and cost analyses reported above.

\textbf{Local models---indoor results:}
Table~\ref{tab:local_indoor} reports the indoor local-model results. In Var.~A, \textit{ministral-3-3b} with CoT reaches the highest room-level accuracy in all three indoor locations, with 66.2\% in the hallway, 72.9\% in the kitchen, and 69.2\% in the office. This suggests that in some models, explicit reasoning can partially compensate for the absence of natural-language threshold comparisons and state-summary flags. However, with enriched prompts, \textit{phi4-mini} becomes the strongest indoor local model. In Var.~C No-CoT mode, it achieves 95.4\%, 97.1\%, and 99.6\% across the hallway, kitchen, and office datasets, respectively. \textit{granite4.1-3b} is also competitive, reaching 86.7--92.1\% in Var.~C No-CoT mode. Local No-CoT latency remains consistently below 0.33\,s per row indoors, whereas CoT increases latency to approximately 2.7--4.9\,s per row.

\begin{table}[!t]
\centering
\footnotesize
\setlength{\tabcolsep}{4pt}
\renewcommand{\arraystretch}{1.12}
\caption{Local-model \textbf{indoor} latency results (hallway, kitchen, office).
         \textit{Acc.} = classification accuracy; \textit{Lat.} = mean latency per row in seconds.
         Bold marks the highest \textit{Acc.} within each dataset and Var.\ column.
         \textbf{C}: \underline{CoT}, and \textbf{N}: \underline{No CoT}.}
\label{tab:local_indoor}
  \resizebox{\columnwidth}{!}{
\begin{tabular}{ll l rr rr rr}
\toprule
\multirow{2}{*}{\rotatebox[origin=c]{90}{\textbf{Dataset}}} & \textbf{Model} & \textbf{Mode} & \multicolumn{2}{c}{\textbf{Var.\,A}} & \multicolumn{2}{c}{\textbf{Var.\,B}} & \multicolumn{2}{c}{\textbf{Var.\,C}} \\
\cmidrule(lr){4-5}\cmidrule(lr){6-7}\cmidrule(lr){8-9}
 & & & \textit{Acc.} & \textit{Lat.} & \textit{Acc.} & \textit{Lat.} & \textit{Acc.} & \textit{Lat.} \\
\midrule
\multirow{10}{*}{\rotatebox[origin=c]{90}{\textbf{Hallway}}} & granite4.1-3b & N & 39.6\% & 0.139 & 80.4\% & 0.160 & 86.7\% & 0.157 \\
 & granite4.1-3b & C & 57.1\% & 4.170 & 75.8\% & 3.741 & 87.5\% & 3.798 \\
 & Llama-3.2-3B & N & 60.0\% & 0.210 & 68.8\% & 0.194 & 84.2\% & 0.209 \\
 & Llama-3.2-3B & C & 55.8\% & 3.017 & 57.9\% & 2.830 & 66.2\% & 2.809 \\
 & llama3.1-8b & N & 47.9\% & 0.326 & 64.6\% & 0.295 & 77.9\% & 0.307 \\
 & llama3.1-8b & C & 46.2\% & 4.824 & 50.4\% & 4.277 & 60.8\% & 4.779 \\
 & ministral-3-3b & N & 50.0\% & 0.218 & 50.0\% & 0.204 & 57.5\% & 0.220 \\
 & ministral-3-3b & C & \textbf{66.2\%} & 2.993 & 80.8\% & 2.806 & 88.3\% & 3.015 \\
 & phi4-mini & N & 50.0\% & 0.220 & \textbf{83.3\%} & 0.207 & \textbf{95.4\%} & 0.222 \\
 & phi4-mini & C & 53.3\% & 3.180 & 63.3\% & 3.488 & 79.2\% & 3.626 \\
\midrule
\multirow{10}{*}{\rotatebox[origin=c]{90}{\textbf{Kitchen}}} & granite4.1-3b & N & 66.2\% & 0.135 & 88.8\% & 0.152 & 92.1\% & 0.157 \\
 & granite4.1-3b & C & 62.9\% & 4.212 & 85.8\% & 3.794 & 92.1\% & 3.845 \\
 & Llama-3.2-3B & N & 55.4\% & 0.198 & 70.0\% & 0.196 & 87.1\% & 0.225 \\
 & Llama-3.2-3B & C & 50.4\% & 3.106 & 64.2\% & 2.751 & 73.8\% & 2.789 \\
 & llama3.1-8b & N & 43.3\% & 0.252 & 70.0\% & 0.290 & 81.2\% & 0.330 \\
 & llama3.1-8b & C & 48.3\% & 4.884 & 55.0\% & 4.325 & 60.8\% & 4.694 \\
 & ministral-3-3b & N & 50.0\% & 0.203 & 50.0\% & 0.213 & 57.9\% & 0.220 \\
 & ministral-3-3b & C & \textbf{72.9\%} & 2.972 & 79.6\% & 2.789 & 89.2\% & 3.047 \\
 & phi4-mini & N & 50.0\% & 0.209 & \textbf{90.4\%} & 0.221 & \textbf{97.1\%} & 0.221 \\
 & phi4-mini & C & 57.1\% & 3.758 & 65.0\% & 3.639 & 93.8\% & 3.591 \\
\midrule
\multirow{10}{*}{\rotatebox[origin=c]{90}{\textbf{Office}}} & granite4.1-3b & N & 45.4\% & 0.141 & 85.4\% & 0.161 & 91.7\% & 0.156 \\
 & granite4.1-3b & C & 55.8\% & 4.253 & 77.1\% & 3.789 & 89.6\% & 3.800 \\
 & Llama-3.2-3B & N & 56.7\% & 0.204 & 68.3\% & 0.196 & 83.8\% & 0.210 \\
 & Llama-3.2-3B & C & 52.9\% & 3.076 & 63.7\% & 2.702 & 69.2\% & 2.795 \\
 & llama3.1-8b & N & 49.2\% & 0.257 & 61.7\% & 0.296 & 74.6\% & 0.312 \\
 & llama3.1-8b & C & 47.1\% & 4.841 & 54.6\% & 4.172 & 57.5\% & 4.661 \\
 & ministral-3-3b & N & 50.0\% & 0.213 & 50.0\% & 0.210 & 58.8\% & 0.222 \\
 & ministral-3-3b & C & \textbf{69.2\%} & 3.009 & 80.0\% & 2.736 & 83.8\% & 3.068 \\
 & phi4-mini & N & 50.0\% & 0.212 & \textbf{91.2\%} & 0.214 & \textbf{99.6\%} & 0.224 \\
 & phi4-mini & C & 59.2\% & 3.729 & 65.0\% & 3.731 & 90.8\% & 3.673 \\
\bottomrule
\end{tabular}
}
\renewcommand{\arraystretch}{1}
\end{table}

\textbf{Local models---outdoor results:}
Table~\ref{tab:local_outdoor} presents the outdoor local-model results. As in the indoor setting, enriched prompts produce the strongest performance. In Var.~B, \textit{granite4.1-3b} is the most stable outdoor model, reaching 91.7\% CoT accuracy in Helsinki, 93.8\% CoT accuracy in Katowice, and 92.5\% accuracy in Warsaw under both inference modes. In Var.~C, \textit{llama3.1-8b} No-CoT achieves the highest outdoor accuracies, with 95.0\% in Helsinki, 97.5\% in Katowice, and 97.9\% in Warsaw. \textit{Llama-3.2-3B} also improves markedly from Var.~A to Var.~C, increasing from 44.6--48.8\% No-CoT accuracy to 91.7--96.2\% No-CoT accuracy across the three outdoor cities. Outdoor local No-CoT latency remains approximately 0.14--0.32\,s per row, while CoT increases latency to approximately 2.6--6.0\,s per row.

\begin{table}[!t]
\centering
\footnotesize
\setlength{\tabcolsep}{4pt}
\renewcommand{\arraystretch}{1.12}
\caption{Local-model \textbf{outdoor} latency results (Helsinki, Katowice, Warsaw).
         \textit{Acc.} = classification accuracy; \textit{Lat.} = mean latency per row in seconds.
         Bold marks the highest \textit{Acc.} within each dataset and Var.\ column.
         \textbf{C}: \underline{CoT}, and \textbf{N}: \underline{No CoT}.}
\label{tab:local_outdoor}
  \resizebox{\columnwidth}{!}{
\begin{tabular}{ll l rr rr rr}
\toprule
\multirow{2}{*}{\rotatebox[origin=c]{90}{\textbf{Dataset}}} & \textbf{Model} & \textbf{Mode} & \multicolumn{2}{c}{\textbf{Var.\,A}} & \multicolumn{2}{c}{\textbf{Var.\,B}} & \multicolumn{2}{c}{\textbf{Var.\,C}} \\
\cmidrule(lr){4-5}\cmidrule(lr){6-7}\cmidrule(lr){8-9}
 & & & \textit{Acc.} & \textit{Lat.} & \textit{Acc.} & \textit{Lat.} & \textit{Acc.} & \textit{Lat.} \\
\midrule
\multirow{10}{*}{\rotatebox[origin=c]{90}{\textbf{Helsinki}}} & granite4.1-3b & N & 76.7\% & 0.141 & 87.5\% & 0.154 & 93.3\% & 0.159 \\
 & granite4.1-3b & C & 67.5\% & 3.394 & \textbf{91.7\%} & 3.170 & 92.1\% & 3.322 \\
 & Llama-3.2-3B & N & 44.6\% & 0.194 & 71.7\% & 0.206 & 91.7\% & 0.210 \\
 & Llama-3.2-3B & C & 63.7\% & 3.152 & 79.6\% & 2.943 & 91.7\% & 2.652 \\
 & llama3.1-8b & N & \textbf{84.2\%} & 0.275 & 85.0\% & 0.291 & \textbf{95.0\%} & 0.306 \\
 & llama3.1-8b & C & 57.9\% & 5.686 & 83.3\% & 4.816 & 91.2\% & 4.962 \\
 & ministral-3-3b & N & 39.2\% & 0.202 & 48.8\% & 0.221 & 61.7\% & 0.226 \\
 & ministral-3-3b & C & 55.8\% & 3.158 & 66.7\% & 2.852 & 72.9\% & 3.043 \\
 & phi4-mini & N & 65.8\% & 0.202 & \textbf{91.7\%} & 0.217 & 92.1\% & 0.223 \\
 & phi4-mini & C & 52.9\% & 4.047 & 76.7\% & 3.771 & 82.9\% & 3.949 \\
\midrule
\multirow{10}{*}{\rotatebox[origin=c]{90}{\textbf{Katowice}}} & granite4.1-3b & N & 76.2\% & 0.143 & 93.3\% & 0.160 & 96.2\% & 0.170 \\
 & granite4.1-3b & C & 69.6\% & 3.485 & \textbf{93.8\%} & 3.218 & 95.0\% & 3.223 \\
 & Llama-3.2-3B & N & 48.8\% & 0.189 & 75.8\% & 0.208 & 96.2\% & 0.215 \\
 & Llama-3.2-3B & C & 68.8\% & 3.339 & 86.7\% & 2.728 & 89.2\% & 2.610 \\
 & llama3.1-8b & N & 67.1\% & 0.285 & 85.8\% & 0.300 & \textbf{97.5\%} & 0.317 \\
 & llama3.1-8b & C & 57.1\% & 5.914 & 70.4\% & 4.704 & 89.2\% & 4.963 \\
 & ministral-3-3b & N & 44.2\% & 0.205 & 56.2\% & 0.225 & 66.2\% & 0.232 \\
 & ministral-3-3b & C & 58.8\% & 3.203 & 65.4\% & 2.826 & 74.2\% & 3.034 \\
 & phi4-mini & N & \textbf{81.2\%} & 0.196 & 92.1\% & 0.223 & 94.2\% & 0.232 \\
 & phi4-mini & C & 59.6\% & 4.844 & 80.4\% & 4.056 & 85.4\% & 4.575 \\
\midrule
\multirow{10}{*}{\rotatebox[origin=c]{90}{\textbf{Warsaw}}} & granite4.1-3b & N & 76.7\% & 0.140 & \textbf{92.5\%} & 0.161 & 95.0\% & 0.163 \\
 & granite4.1-3b & C & 74.2\% & 3.558 & \textbf{92.5\%} & 3.200 & 94.2\% & 3.344 \\
 & Llama-3.2-3B & N & 47.9\% & 0.188 & 80.0\% & 0.202 & 94.6\% & 0.214 \\
 & Llama-3.2-3B & C & 73.3\% & 3.363 & 88.8\% & 2.774 & 89.2\% & 2.598 \\
 & llama3.1-8b & N & 70.4\% & 0.274 & 91.7\% & 0.296 & \textbf{97.9\%} & 0.310 \\
 & llama3.1-8b & C & 60.4\% & 5.977 & 77.5\% & 4.767 & 88.3\% & 5.019 \\
 & ministral-3-3b & N & 43.3\% & 0.197 & 62.5\% & 0.225 & 75.4\% & 0.229 \\
 & ministral-3-3b & C & 57.9\% & 3.285 & 70.8\% & 5.621 & 80.8\% & 3.126 \\
 & phi4-mini & N & \textbf{89.6\%} & 0.199 & 91.7\% & 0.225 & 92.5\% & 0.227 \\
 & phi4-mini & C & 64.6\% & 4.796 & 78.8\% & 4.111 & 86.2\% & 4.461 \\
\bottomrule
\end{tabular}
}
\renewcommand{\arraystretch}{1}
\end{table}

\textbf{Cloud models---indoor results:}
Table~\ref{tab:cloud_indoor} reports the indoor cloud-model results. \textit{gemini-2.5-flash} is the strongest indoor cloud model overall, achieving 98.3--100.0\% in Var.~C across rooms and inference modes. \textit{gpt-5.4-mini} is competitive in No-CoT mode, reaching 97.5\% in the kitchen and 98.3\% in the office dataset under Var.~C. However, its CoT accuracy drops substantially across indoor datasets, producing the largest indoor CoT penalty in the benchmark. Cloud No-CoT latency ranges from below 1.2\,s for the faster GPT-family models to approximately 2.9--3.3\,s for \textit{gemini-2.5-flash}.

\begin{table}[!t]
\centering
\footnotesize
\setlength{\tabcolsep}{4pt}
\renewcommand{\arraystretch}{1.12}
\caption{Cloud-model \textbf{indoor} latency results (hallway, kitchen, office).
         \textit{Acc.} = classification accuracy; \textit{Lat.} = mean latency per row in seconds.
         Bold marks the highest \textit{Acc.} within each dataset and Var.\ column.
         \textbf{C}: \underline{CoT}, and \textbf{N}: \underline{No CoT}.}
\label{tab:cloud_indoor}
  \resizebox{\columnwidth}{!}{
\begin{tabular}{ll l rr rr rr}
\toprule
\multirow{2}{*}{\rotatebox[origin=c]{90}{\textbf{Dataset}}} & \textbf{Model} & \textbf{Mode} & \multicolumn{2}{c}{\textbf{Var.\,A}} & \multicolumn{2}{c}{\textbf{Var.\,B}} & \multicolumn{2}{c}{\textbf{Var.\,C}} \\
\cmidrule(lr){4-5}\cmidrule(lr){6-7}\cmidrule(lr){8-9}
 & & & \textit{Acc.} & \textit{Lat.} & \textit{Acc.} & \textit{Lat.} & \textit{Acc.} & \textit{Lat.} \\
\midrule
\multirow{10}{*}{\rotatebox[origin=c]{90}{\textbf{Hallway}}} & gemini-2.5-flash & N & \textbf{88.8\%} & 3.045 & \textbf{85.4\%} & 3.201 & 98.3\% & 2.887 \\
 & gemini-2.5-flash & C & 82.9\% & 3.790 & \textbf{85.4\%} & 3.498 & \textbf{98.8\%} & 3.715 \\
 & gemini-3.1-flash-lite & N & 72.1\% & 1.306 & 66.7\% & 2.110 & 90.4\% & 2.128 \\
 & gemini-3.1-flash-lite & C & 67.9\% & 1.751 & 60.4\% & 1.749 & 84.6\% & 2.310 \\
 & gpt-5.4 & N & 66.7\% & 0.707 & 58.8\% & 0.757 & 77.5\% & 0.781 \\
 & gpt-5.4 & C & 57.1\% & 3.921 & 62.9\% & 3.209 & 70.4\% & 3.061 \\
 & gpt-5.4-mini & N & 85.4\% & 0.665 & 82.1\% & 0.613 & 91.7\% & 0.678 \\
 & gpt-5.4-mini & C & 48.3\% & 2.097 & 47.5\% & 2.093 & 63.3\% & 1.641 \\
 & gpt-5.4-nano & N & 48.3\% & 0.648 & 52.5\% & 0.707 & 82.9\% & 0.740 \\
 & gpt-5.4-nano & C & 48.3\% & 3.016 & 51.7\% & 2.936 & 65.8\% & 2.538 \\
\midrule
\multirow{10}{*}{\rotatebox[origin=c]{90}{\textbf{Kitchen}}} & gemini-2.5-flash & N & \textbf{95.4\%} & 3.075 & \textbf{90.8\%} & 3.094 & \textbf{99.6\%} & 3.083 \\
 & gemini-2.5-flash & C & 90.4\% & 3.654 & 90.4\% & 3.477 & 99.2\% & 3.598 \\
 & gemini-3.1-flash-lite & N & 74.6\% & 1.347 & 81.2\% & 2.097 & 96.7\% & 2.098 \\
 & gemini-3.1-flash-lite & C & 68.3\% & 2.561 & 70.0\% & 1.749 & 87.1\% & 2.502 \\
 & gpt-5.4 & N & 70.0\% & 0.763 & 65.4\% & 0.699 & 72.5\% & 0.684 \\
 & gpt-5.4 & C & 57.5\% & 4.052 & 63.3\% & 2.823 & 75.4\% & 2.999 \\
 & gpt-5.4-mini & N & 86.7\% & 0.671 & 89.6\% & 0.601 & 97.5\% & 0.566 \\
 & gpt-5.4-mini & C & 48.3\% & 1.843 & 48.3\% & 1.791 & 62.5\% & 1.761 \\
 & gpt-5.4-nano & N & 49.2\% & 0.673 & 60.0\% & 0.704 & 80.0\% & 0.674 \\
 & gpt-5.4-nano & C & 50.0\% & 2.646 & 51.2\% & 2.742 & 64.2\% & 2.467 \\
\midrule
\multirow{10}{*}{\rotatebox[origin=c]{90}{\textbf{Office}}} & gemini-2.5-flash & N & \textbf{94.6\%} & 3.155 & 90.4\% & 3.315 & 99.6\% & 2.898 \\
 & gemini-2.5-flash & C & 87.1\% & 3.840 & 89.2\% & 3.492 & \textbf{100.0\%} & 3.612 \\
 & gemini-3.1-flash-lite & N & 77.9\% & 1.351 & 77.9\% & 2.092 & 95.8\% & 2.131 \\
 & gemini-3.1-flash-lite & C & 68.3\% & 1.747 & 67.1\% & 1.693 & 83.8\% & 2.493 \\
 & gpt-5.4 & N & 69.6\% & 0.622 & 64.6\% & 0.657 & 74.2\% & 1.129 \\
 & gpt-5.4 & C & 60.4\% & 3.322 & 59.2\% & 3.314 & 73.3\% & 3.162 \\
 & gpt-5.4-mini & N & 89.6\% & 0.717 & \textbf{92.9\%} & 0.628 & 98.3\% & 0.752 \\
 & gpt-5.4-mini & C & 46.2\% & 2.072 & 47.1\% & 1.775 & 61.7\% & 1.743 \\
 & gpt-5.4-nano & N & 47.9\% & 0.647 & 56.2\% & 0.710 & 74.6\% & 0.649 \\
 & gpt-5.4-nano & C & 49.6\% & 2.622 & 50.4\% & 2.916 & 59.6\% & 2.757 \\
\bottomrule
\end{tabular}
}
\renewcommand{\arraystretch}{1}
\end{table}

\textbf{Cloud models---outdoor results:}
Table~\ref{tab:cloud_outdoor} presents the outdoor cloud-model results. Outdoor cloud performance is strong across almost all evaluated models. \textit{gemini-3.1-flash-lite} is the most consistently accurate cloud model in Var.~C No-CoT mode, reaching 97.5--98.8\%  across all three datasets. \textit{gemini-2.5-flash} achieves the highest peak results of 99.6--100.0\% in Var.~C CoT mode. Among GPT-family models, \textit{gpt-5.4} is the strongest in No-CoT mode, reaching 87.9--97.9\% across Var.~A--Var.~C and the three datasets. GPT-family models are faster than those from the Gemini family, remaining below 1\,s No-CoT latency per row.

\begin{table}[!t]
\centering
\footnotesize
\setlength{\tabcolsep}{4pt}
\renewcommand{\arraystretch}{1.12}
\caption{Cloud-model \textbf{outdoor} latency results (Helsinki, Katowice, Warsaw).
         \textit{Acc.} = classification accuracy; \textit{Lat.} = mean latency per row in seconds.
         Bold marks the highest \textit{Acc.} within each dataset and Var.\ column.
         \textbf{C}: \underline{CoT}, and \textbf{N}: \underline{No CoT}.}
\label{tab:cloud_outdoor}
  \resizebox{\columnwidth}{!}{
\begin{tabular}{ll l rr rr rr}
\toprule
\multirow{2}{*}{\rotatebox[origin=c]{90}{\textbf{Dataset}}} & \textbf{Model} & \textbf{Mode} & \multicolumn{2}{c}{\textbf{Var.\,A}} & \multicolumn{2}{c}{\textbf{Var.\,B}} & \multicolumn{2}{c}{\textbf{Var.\,C}} \\
\cmidrule(lr){4-5}\cmidrule(lr){6-7}\cmidrule(lr){8-9}
 & & & \textit{Acc.} & \textit{Lat.} & \textit{Acc.} & \textit{Lat.} & \textit{Acc.} & \textit{Lat.} \\
\midrule
\multirow{10}{*}{\rotatebox[origin=c]{90}{\textbf{Helsinki}}} & gemini-2.5-flash & N & 72.9\% & 4.100 & 60.8\% & 4.193 & 89.2\% & 3.482 \\
 & gemini-2.5-flash & C & 91.7\% & 4.094 & 74.2\% & 4.296 & \textbf{99.6\%} & 3.828 \\
 & gemini-3.1-flash-lite & N & \textbf{96.2\%} & 1.547 & \textbf{96.2\%} & 2.091 & 97.5\% & 2.092 \\
 & gemini-3.1-flash-lite & C & 94.6\% & 2.006 & 95.8\% & 1.661 & 96.2\% & 1.764 \\
 & gpt-5.4 & N & 90.0\% & 0.902 & 87.9\% & 0.897 & 97.1\% & 0.675 \\
 & gpt-5.4 & C & 90.0\% & 3.045 & 89.6\% & 2.789 & 92.1\% & 2.767 \\
 & gpt-5.4-mini & N & 86.2\% & 0.671 & 85.0\% & 0.692 & 92.9\% & 0.615 \\
 & gpt-5.4-mini & C & 88.3\% & 1.953 & 85.8\% & 1.656 & 89.6\% & 1.669 \\
 & gpt-5.4-nano & N & 83.3\% & 0.857 & 85.0\% & 0.650 & 86.2\% & 0.817 \\
 & gpt-5.4-nano & C & 80.4\% & 2.126 & 66.2\% & 2.213 & 90.8\% & 2.198 \\
\midrule
\multirow{10}{*}{\rotatebox[origin=c]{90}{\textbf{Katowice}}} & gemini-2.5-flash & N & 95.4\% & 3.284 & 92.5\% & 2.874 & 97.9\% & 3.058 \\
 & gemini-2.5-flash & C & 95.8\% & 3.820 & 94.2\% & 3.302 & \textbf{100.0\%} & 3.499 \\
 & gemini-3.1-flash-lite & N & 94.6\% & 1.329 & 95.4\% & 2.104 & 98.8\% & 2.185 \\
 & gemini-3.1-flash-lite & C & \textbf{97.9\%} & 2.487 & \textbf{97.5\%} & 1.661 & 98.8\% & 1.817 \\
 & gpt-5.4 & N & 95.8\% & 0.801 & 96.2\% & 0.752 & 97.1\% & 0.789 \\
 & gpt-5.4 & C & 93.3\% & 2.979 & 92.9\% & 2.751 & 95.4\% & 2.578 \\
 & gpt-5.4-mini & N & 89.2\% & 0.772 & 92.9\% & 0.725 & 96.2\% & 0.662 \\
 & gpt-5.4-mini & C & 90.0\% & 1.896 & 87.5\% & 1.643 & 92.9\% & 1.710 \\
 & gpt-5.4-nano & N & 86.7\% & 0.609 & 87.9\% & 0.630 & 90.4\% & 0.601 \\
 & gpt-5.4-nano & C & 89.2\% & 1.989 & 88.3\% & 2.353 & 96.2\% & 2.360 \\
\midrule
\multirow{10}{*}{\rotatebox[origin=c]{90}{\textbf{Warsaw}}} & gemini-2.5-flash & N & 97.5\% & 3.283 & 96.7\% & 3.138 & 99.6\% & 2.706 \\
 & gemini-2.5-flash & C & \textbf{99.6\%} & 3.650 & 96.2\% & 3.477 & \textbf{100.0\%} & 3.469 \\
 & gemini-3.1-flash-lite & N & 95.8\% & 2.116 & 95.0\% & 2.119 & 97.9\% & 1.602 \\
 & gemini-3.1-flash-lite & C & 96.7\% & 1.683 & \textbf{97.5\%} & 1.718 & 98.3\% & 2.429 \\
 & gpt-5.4 & N & 96.7\% & 0.800 & 97.1\% & 0.691 & 97.9\% & 0.727 \\
 & gpt-5.4 & C & 92.1\% & 2.485 & 91.2\% & 2.282 & 93.8\% & 2.223 \\
 & gpt-5.4-mini & N & 88.3\% & 0.680 & 91.2\% & 0.600 & 94.2\% & 0.621 \\
 & gpt-5.4-mini & C & 90.8\% & 1.448 & 85.8\% & 1.228 & 89.6\% & 1.248 \\
 & gpt-5.4-nano & N & 86.2\% & 0.633 & 86.2\% & 0.695 & 89.6\% & 0.630 \\
 & gpt-5.4-nano & C & 89.2\% & 1.904 & 91.2\% & 2.031 & 97.1\% & 2.047 \\
\bottomrule
\end{tabular}
}
\renewcommand{\arraystretch}{1}
\end{table}

\subsection{Summary of Results}\label{sec:summary}

Five synthesis points emerge from the aggregate, error, and cost analyses. First, prompt enrichment (Var.~A$\to$Var.~C) is the dominant performance driver: the cumulative $\Delta_{A\to C} = +30.8$\,pp for local indoor No-CoT exceeds the contribution of model scale, inference mode, or the local-vs-cloud boundary, and the accuracy gain is essentially free in compute terms since output token count remains at $\approx$1 across all prompt variants. Second, under Var.~C No-CoT, the best local models (\textit{phi4-mini}, on average, 97.4\% -- indoor; \textit{llama3.1-8b}, 96.8\% -- outdoor) are within 0--2.5\,pp of the best cloud models, with operational costs even a few hundred times lower. Third, CoT is conditionally useful only in the raw-prompt Var.~A setting for local models; in enriched conditions, it universally degrades cloud indoor accuracy (mean penalty $-12$\,pp; worst case $-37.8$\,pp for \textit{gpt-5.4-mini}) and introduces reasoning failures and parse errors in local models without accuracy benefit. Fourth, latency follows a four-level hierarchy (local No-CoT $\approx$0.22\,s $<$ cloud No-CoT $\approx$1.4\,s $<$ cloud CoT $\approx$2.5\,s $<$ local CoT $\approx$3.7\,s): only Var.~C No-CoT achieves both near-cloud accuracy and sub-0.3\,s latency simultaneously. Fifth, the primary CoT failure mode --- answer--reasoning inversion driven by training-prior polarity --- is entirely absent in No-CoT mode, strengthening the case for prompt enrichment as the correct engineering lever over inference-time reasoning for reliable IoT sensor diagnostics.

\section{Discussion}\label{sec:Discussion}

The main observation is that local models benefit substantially when raw measurements are transformed into threshold-aware and semantically enriched textual representations. This indicates that the performance gap between local and cloud models is not only a matter of model size, but also a matter of how sensor information is represented before inference. In this sense, the proposed preprocessing strategy acts as an interface layer between numerical IoT data and language-model reasoning. This extends the expected conclusion that prompt construction plays a central role in enabling local LLMs to perform air-quality and thermal-comfort classification from IoT sensor data. 

\textbf{Effect of prompt enrichment:}
Variant A provides only raw sensor values and the user query. This setting requires the model to infer, reconstruct, or approximate the task-relevant meaning of each measurement and its acceptable range from an underspecified textual representation. This baseline setting is consistent with prior work on prompt-based learning, where the input template determines how a pretrained language model can map an input instance to a prediction~\cite{Liu2023Prompting}. As a result of the gap between raw numerical readings and their task-relevant interpretation, local models obtain the weakest performance in this configuration. Variant B addresses this gap by adding threshold-aware information, allowing the model to compare each sensor value directly with its corresponding acceptable range. This finding is consistent with recent work on IoT-oriented LLM reasoning, where raw sensor values are reported to be difficult for LLMs to interpret unless they are accompanied by additional contextual information that clarifies their physical and task-specific meaning~\cite{An2026IoT}. This produces the largest improvement for local models, suggesting that explicit environmental limits are more useful than relying on the model's implicit domain knowledge when the model must operate without cloud-scale inference support. This finding also complements recent IoT-oriented LLM research, which emphasizes the need to bridge heterogeneous sensor streams and language-model reasoning \cite{Mo2024IoTLM}, while showing that, for this binary environmental-assessment task, a lightweight prompt-level bridge is enough to substantially reduce the gap between local and cloud models. Variant C further adds compact meta-information, such as an air-quality flag, which provides an additional improvement in many local-model settings. However, the gain from Var.~B to Var.~C is smaller than the gain from Var.~A to Var.~B, indicating that threshold information is the most important component of the proposed prompt enrichment strategy. From the perspective of edge and IoT deployment, this is important because local inference is often motivated by latency, privacy, cost, and connectivity constraints \cite{DeVito2025IoT}; our results indicate that part of the performance normally attributed to larger cloud models can instead be recovered by a better representation of sensor data before inference.

\textbf{Chain-of-thought:}
The results also show that chain-of-thought should not be used as a default inference strategy for this task. Although CoT prompting has been shown to improve performance on complex arithmetic, commonsense, and symbolic reasoning tasks \cite{wei2022chain}, its benefits should be treated as task-dependent rather than automatic. In the present experiments, CoT is useful mainly when the prompt is weakly contextualized, especially in Variant A, where the model receives only raw sensor values. In that setting, reasoning-style prompting may help some local models compensate for the lack of explicit threshold information. However, once threshold-aware information is provided in Variant B and Variant C, CoT often reduces reliability or accuracy. This is consistent with recent evidence that inference-time reasoning can also reduce model performance when additional verbal deliberation introduces unnecessary or misleading intermediate steps \cite{Liu2024Mind}. In our setting, the likely issue is not that reasoning is useless, but that the task requires a short rule-based classification, while CoT encourages a longer free-form response. This distinction is particularly important for deployment because the system requires short, machine-readable binary outputs. A correct-looking reasoning trace is less useful if the final answer contradicts it, cannot be parsed reliably, or fails to follow the required output format \cite{Turpin2023Language,Molfese2025Right}. Therefore, No-CoT prompting is more suitable as the default mode for sensor-network deployment, while CoT should be reserved for cases where explanation is required and output parsing can be controlled. If explanation is necessary, techniques such as answer-consistency checks or self-consistency decoding may be combined with CoT, but they introduce additional inference cost and are therefore less attractive for disconnected or latency-sensitive operation \cite{Wang2022SelfConsistency}.

\textbf{Model size and model selection:}
The results suggest that model size alone is not a sufficient criterion for selecting a local model. Scaling laws show that language-model performance generally improves with model size, data, and training compute \cite{Kaplan2020Scaling}, but later work on compute-optimal training also shows that parameter count cannot be interpreted independently from the amount and quality of training data \cite{hoffmann2022training}. For deployment in smart environments, this distinction is essential: a larger local model may have greater representational capacity, but its practical performance also depends on architecture, instruction tuning, quantization, and compatibility with the prompt format. Prior work on instruction following has shown that smaller instruction-tuned models can be preferred over much larger base models when the goal is to follow user intent reliably \cite{Ouyang2022Training}. Similarly, recent efficient open models demonstrate that architectural choices and training recipes can allow smaller models to outperform larger alternatives on several benchmarks \cite{Jiang2023Mistral}. Our results extend this observation to sensor-based smart-environment classification: model selection for edge AI should be task-specific rather than based only on the number of parameters. In practical deployments, a smaller model with strong instruction-following ability, stable behavior under threshold-aware prompts, and robust output parsing may be preferable to a larger model that is less reliable in the required binary decision format. This is especially relevant when local deployment relies on compression or quantization, since the same model may behave differently depending on the inference setup \cite{Dettmers2022LLM}.

\textbf{Edge--cloud deployment implications:}
Cloud models achieve strong performance under direct No-CoT prompting, confirming their value as high-accuracy reference models. However, the results also show that cloud inference does not automatically benefit from additional reasoning instructions. In several cases, CoT reduces parse reliability, especially when the model produces longer responses instead of the requested binary answer. Taken together with the strong performance of enriched local prompts, these results show that routine smart-environment classification does not necessarily require cloud inference. A more appropriate deployment view is therefore functional role allocation: local models handle routine decisions, e.g. air-quality and comfort, when the prompt contains threshold and summary information, while cloud models are reserved for uncertain cases, more complex queries, or tasks requiring richer explanations. This division of roles is aligned with recent work on LLM-enabled edge AI and cloud--edge collaboration, where local or edge-side models provide low-latency operation and cloud models are used selectively when higher reasoning capacity is needed \cite{Shen2024Large,Jin2024CE}. Such a design is suitable for smart environments because it reduces dependence on cloud connectivity while preserving the option of cloud-level reasoning when the network is available. It also addresses a practical challenge identified in LLM-for-IoT research: LLMs can enrich IoT systems, but their computational cost, privacy implications, and deployment complexity require careful integration rather than unconditional cloud reliance~\cite{DeVito2025IoT}.

\textbf{Implications for IoT analytics:}
The proposed methodology reframes part of IoT preprocessing as a prompt-level knowledge-integration problem. Instead of assuming the need to fine-tune every local model or retrain a dedicated classifier for each sensing context, the system injects domain knowledge through structured textual descriptions. This idea is consistent with earlier semantic-sensor research, which argued that sensor observations become more useful when raw readings are accompanied by contextual metadata and machine-interpretable descriptions \cite{Sheth2008Semantic,Janowicz2019SOSA}. In our case, this contextualization is performed not by a separate semantic layer alone, but by the prompt itself: limits, threshold violations, and summary flags are written directly into the model input. This is particularly relevant for smart environments, where sensor types, acceptable ranges, and user requirements may change over time. The approach also follows the broader logic of prompt-based learning, in which a model is adapted to a task by changing the input template rather than updating model parameters \cite{Liu2023Prompting}. Recent IoT-oriented LLM research similarly shows that raw sensor values are difficult for LLMs to interpret unless they are transformed into representations enriched with task-relevant knowledge \cite{An2026IoT}. Our results extend this direction by showing that such enrichment can be sufficient for local models to perform routine environmental classification with high reliability. Finally, because the relevant thresholds and rules are explicit in the prompt, the system is more transparent and easier to update than a model in which the same information is hidden inside learned parameters. In future deployments, this prompt-level strategy could also be combined with retrieval-augmented generation, where current standards, room-specific policies, or user preferences are retrieved dynamically and inserted into the prompt \cite{Lewis2020Retrieval}.

\textbf{Limitations and future work:}
The current evaluation focuses on binary yes/no classification tasks. Although this simplifies evaluation and parsing, future work should examine multi-class decisions, open-ended explanations, and temporal reasoning over longer sensor histories. The thresholds used in the prompts are also domain-dependent; therefore, future systems should support dynamic retrieval of authoritative standards, such as air-quality regulations and building-comfort guidelines, through a controlled RAG component. This would follow the broader idea of retrieval-augmented generation, where the model is supported by an explicit external knowledge source rather than relying only on information stored in its parameters \cite{Lewis2020Retrieval}. Another limitation is that the present evaluation emphasizes accuracy, parse rate, and latency, while practical edge deployment also requires measurements of memory consumption, energy use, and robustness under missing or noisy sensor readings. These factors are central in LLM-for-IoT and edge-LLM deployment, where computation, privacy, hardware constraints, and integration complexity remain major practical challenges \cite{DeVito2025IoT,Zheng2025Review}. Finally, future experiments should evaluate the full pipeline on resource-constrained hardware, such as Raspberry Pi devices, to verify whether the observed accuracy gains remain practical under real edge-computing constraints.

\section{Conclusion}\label{sec:Conclusion}

This paper investigated whether prompt-side preprocessing can improve the ability of locally deployed LLMs to interpret IoT sensor data for air-quality and thermal-comfort classification. We proposed a structured prompt construction framework comprising three progressively enriched variants --- raw sensor values, threshold-aware descriptions, and compact environmental summary flags --- and evaluated them across five local and five cloud LLMs on indoor and outdoor datasets under both No-CoT and CoT inference modes. The results establish prompt enrichment as the dominant performance driver, yielding cumulative accuracy gains of $+30.8$\,pp for local indoor No-CoT and $+25.6$\,pp for local outdoor No-CoT inference, and demonstrate that the performance gap between local and cloud deployment is substantially a representation problem rather than a model-capacity problem: under Variant~C No-CoT, \textit{phi4-mini} reaches $97.4\%$ indoor accuracy and \textit{llama3.1-8b} reaches $96.8$--$97.9\%$ across outdoor datasets --- within $0$--$2.5$\,pp of the best cloud models, followed by up to $210\times$ lower operational cost, sub-$0.3$\,s latency, and zero parse failures, establishing it as the optimal operating point for latency-sensitive and privacy-aware IoT deployments. The analysis further demonstrates that chain-of-thought prompting should not serve as a default inference strategy for structured IoT classification: once threshold-aware context is provided, CoT universally degrades cloud indoor accuracy by a mean of $-12$\,pp, with a worst case of $-37.8$\,pp for \textit{gpt-5.4-mini}, introduces answer--reasoning inversions and parse failures in local models, and offers no measurable accuracy benefit in any enriched-prompt configuration, confirming that structured prompt-level knowledge integration is a more effective and resource-efficient strategy than either model scaling or CoT prompting --- a principle likely to generalise beyond environmental monitoring to other threshold-driven IoT classification tasks where compact edge models must operate reliably under real-world resource constraints.

\section*{Acknowledgments}
This research was partially supported by the Jane and Aatos Erkko Foundation EVIL-AI project, the Ministry of National Education of the Republic of Türkiye, and the Research Council of Finland with grant number 362594.

This work was also partially supported by the Reactive Too project that has received funding from the European Union’s Horizon 2020 Research, Innovation and Staff Exchange Programme under the Marie Sklodowska-Curie Action (Grant Agreement No 871163). Scientific work published as part of an international project ReACTIVE Too was co-financed by the program of the Minister of Science and Higher Education entitled “PMW” in the years 2024–2025; contract no. 5872/H2020/2024/2.

\bibliographystyle{IEEEtran}
\bibliography{References}

\vspace{-25pt}

\begin{IEEEbiography}[{\includegraphics[width=1in,height=1.25in,clip,keepaspectratio]{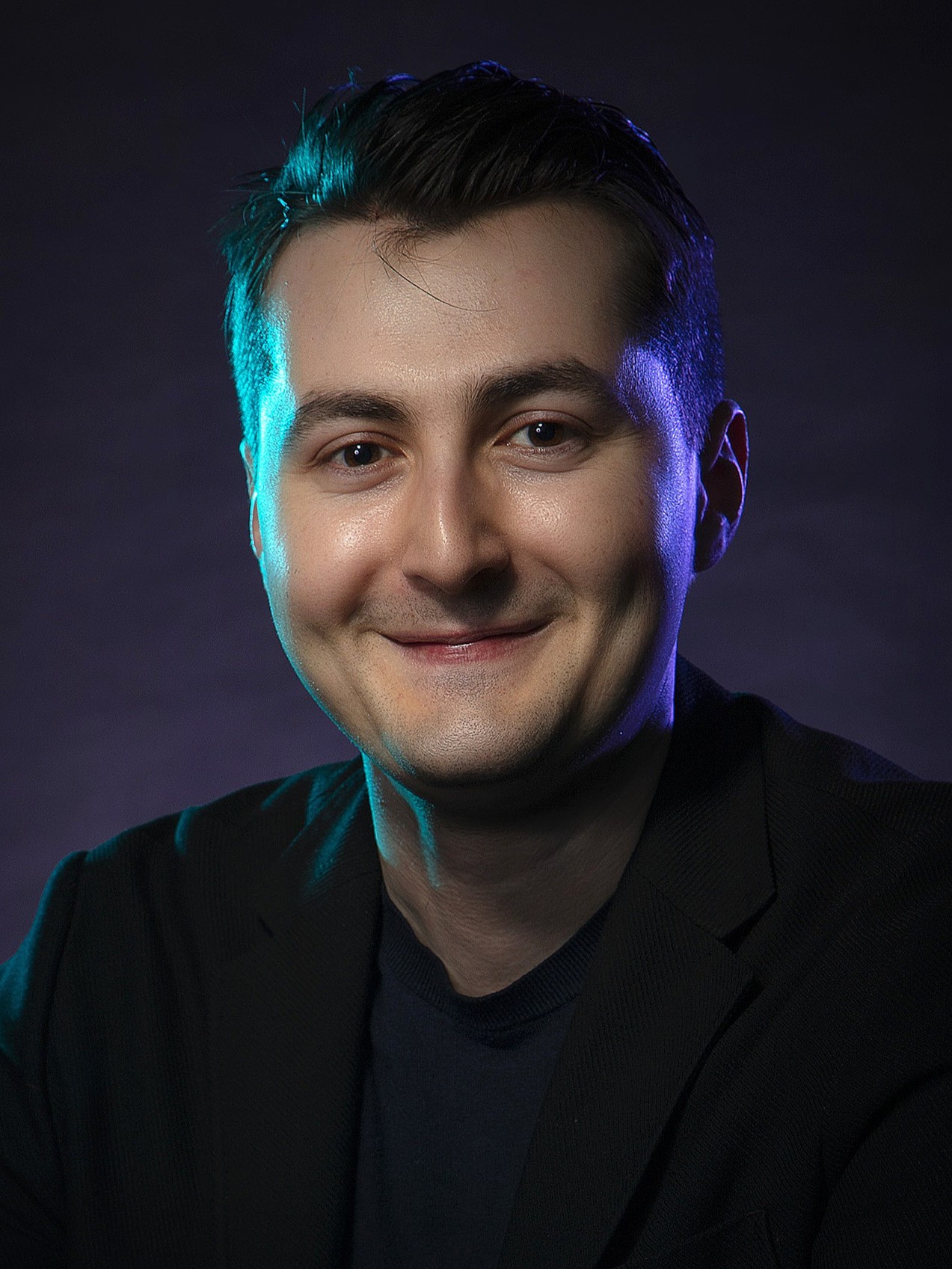}}]{Ayg\"{u}n Varol}
received the M.Sc. degree in electrical and electronics engineering from the Isparta University of Applied Sciences, Isparta, T\"{u}rkiye, in 2022. He is currently a doctoral candidate with the Faculty of Information Technology and Communication Sciences, Tampere University, Tampere, Finland. His research interests include the Internet of Things, smart environments, artificial intelligence and machine learning (AI/ML), and the identification and mitigation of the negative impacts of AI. 
\end{IEEEbiography}

\vspace{-25pt}

\begin{IEEEbiography}[{\includegraphics[width=1in,height=1.25in,clip,keepaspectratio, angle=-90]{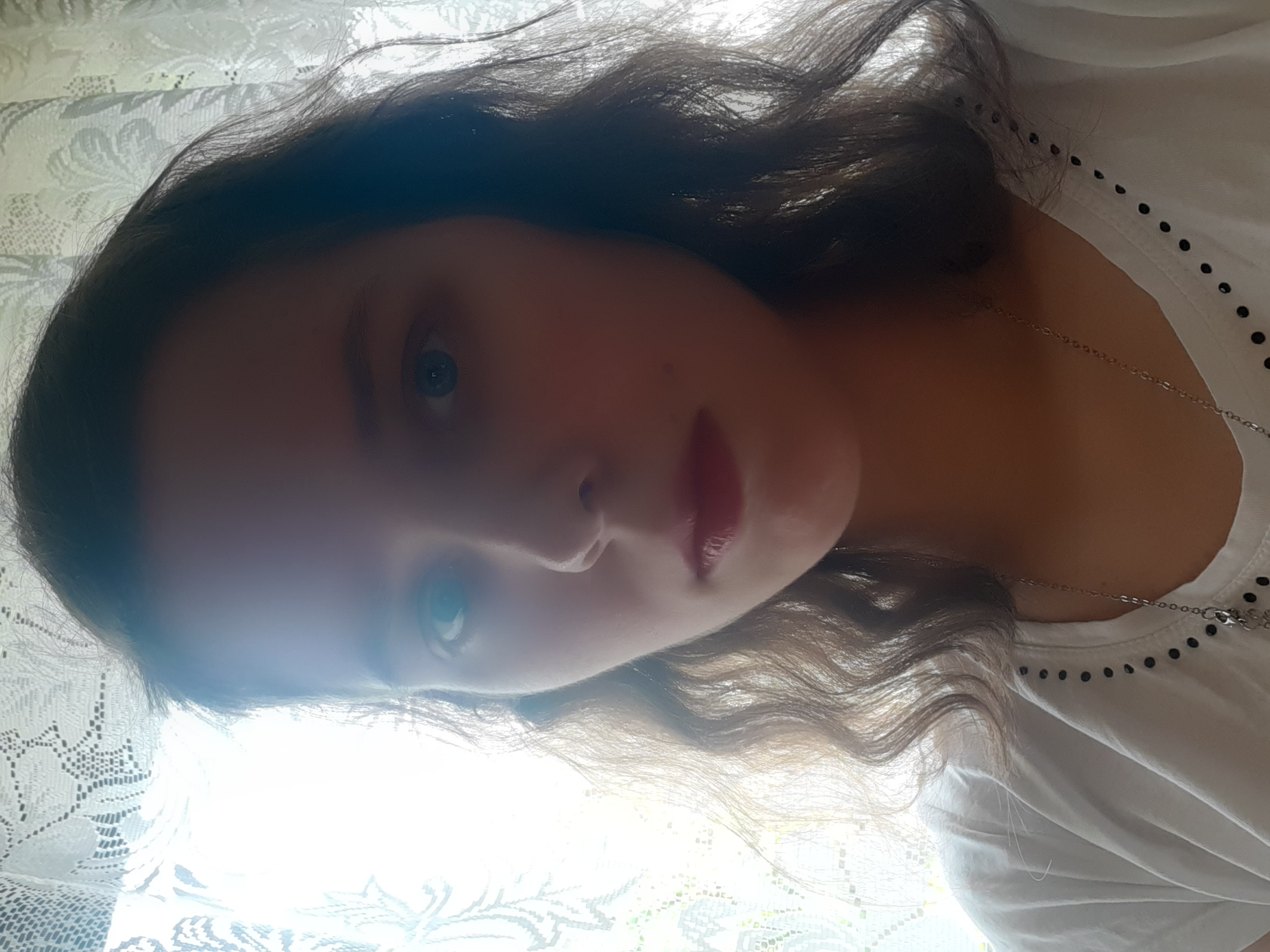}}]{Katarzyna Ko{\l}odziej}
received the M.Sc. degree in power engineering from the Faculty of Energy and Environmental Engineering, Silesian University of Technology, Gliwice, Poland, in 2021. Since 2022, she has been an Assistant with the Machine Learning Team, Institute of Theoretical and Applied Informatics, Polish Academy of Sciences (ITAI PAS), Gliwice, Poland. Her research interests include large language models, large multimodal models, computer vision, and AI applications for solving engineering problems.
\end{IEEEbiography}

\vspace{-25pt}

\begin{IEEEbiography}[{\includegraphics[width=1in,height=1.25in,clip,keepaspectratio]{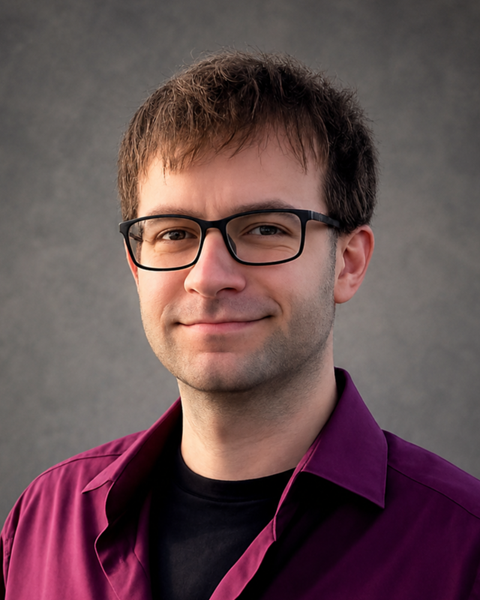}}]{{\L}ukasz Sobczak}
obtained his B.Sc. and M.Sc degrees in Computer Science at the Silesian University of Technology (Gliwice, Poland) in 2016 and 2017. He participated in the research project "Modular Autonomy System with a Virtual Testing Environment" supported by the program "Future Technologies for Defense - Young Scientists Competition 2017" of NCBR (Polish National Center for Research and Development). He received his Ph.D. degree in Computer Science from the same university in 2023.
His current research interests include robotics applications, autonomous vehicles, large language models and IoT.
\end{IEEEbiography}

\vspace{-25pt}

\begin{IEEEbiography}[{\includegraphics[width=1in,height=1.25in,clip,keepaspectratio]{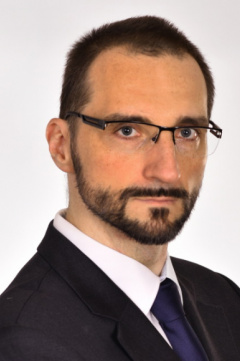}}]{Micha{\l} Romaszewski}
obtained Ph.D. degree in computer science from the Institute of Theoretical and Applied Informatics, Polish Academy of Sciences (ITAI PAS), Gliwice, Poland. He is currently an Assistant Professor with the Machine Learning Team, ITAI PAS. His research focuses on large language models, hyperspectral data analysis and applied machine learning. 
\end{IEEEbiography}

\vspace{-25pt}

\begin{IEEEbiography}[{\includegraphics[width=1in,height=1.25in,clip,keepaspectratio]{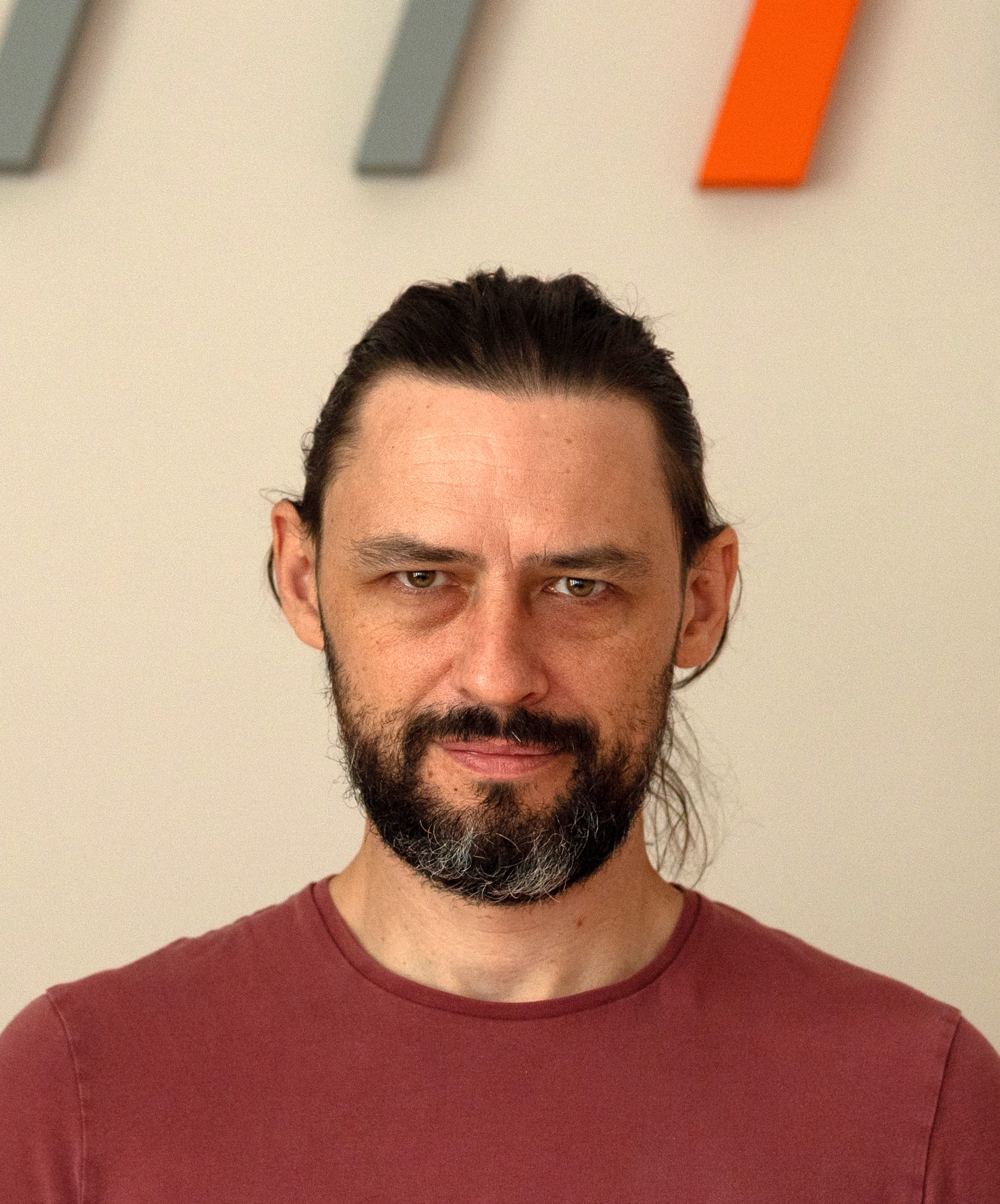}}]{Przemys{\l}aw G{\l}omb} received the Ph.D. degree in informatics from the Institute of Theoretical and Applied Informatics, Polish Academy of Sciences (ITAI PAS), Gliwice, Poland, in 2006, and the habilitation degree in technical informatics and telecommunications from the Silesian University of Technology, Gliwice, Poland, in 2020. He is currently a Professor of the Institute and the Head of the Machine Learning Group at ITAI PAS. His research interests span theoretical and practical problems in machine learning, including the design of neural architectures for time-series and computer-vision problems, bridging signal processing and large language models, and adapting machine learning models for industrial, cultural, and health-related applications.
\end{IEEEbiography}

\vspace{-25pt}

\begin{IEEEbiography}[{\includegraphics[width=1in,height=1.25in,clip,keepaspectratio]{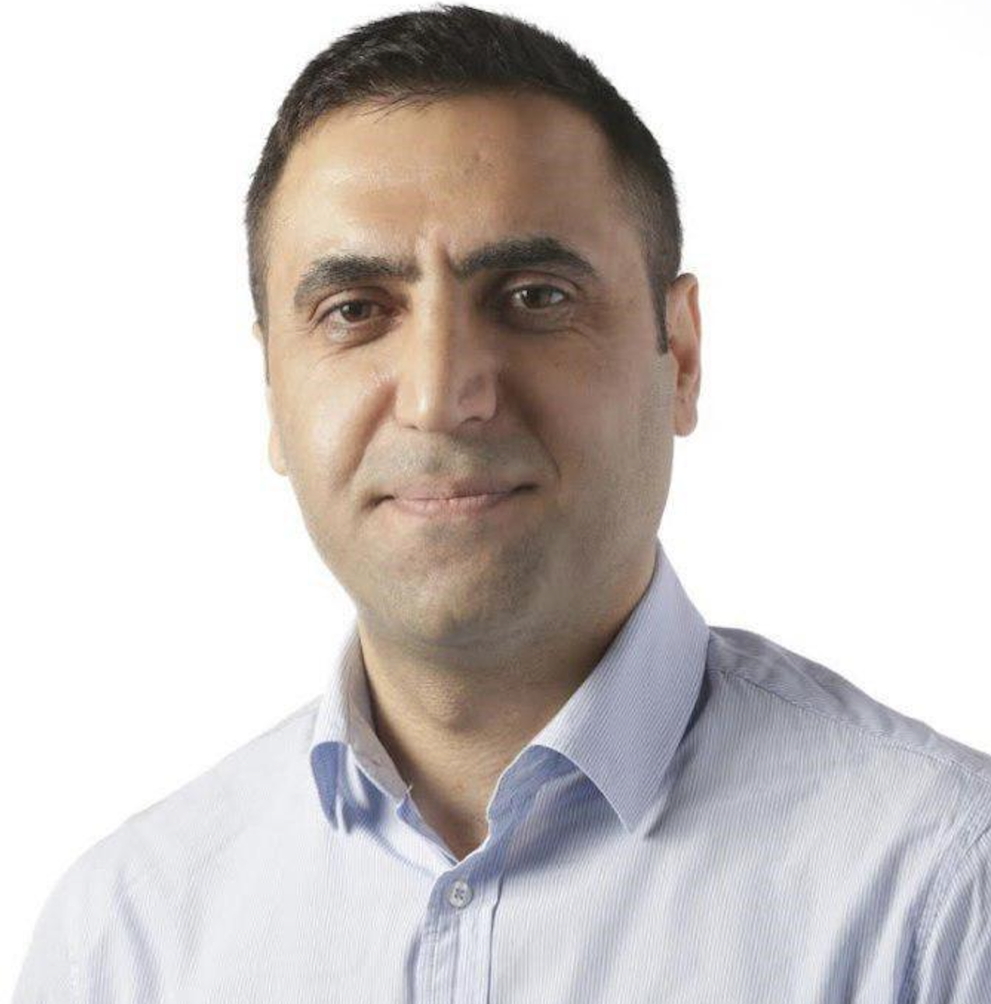}}]{Naser~Hossein~Motlagh}
received the D.Sc. degree in networking technology from the School of Electrical Engineering, Aalto University, Espoo, Finland, in 2018. He is a Senior Researcher with the Department of Computer Science, University of Helsinki, Helsinki, Finland, within the Nokia Center for Advanced Research (NCAR). His research interests include the Internet of Things, wireless sensor networks, environmental sensing, smart buildings, and unmanned aerial and underwater vehicles.
\end{IEEEbiography}

\vspace{-25pt}

\begin{IEEEbiography}[{\includegraphics[width=1in,height=1.25in,clip,keepaspectratio]{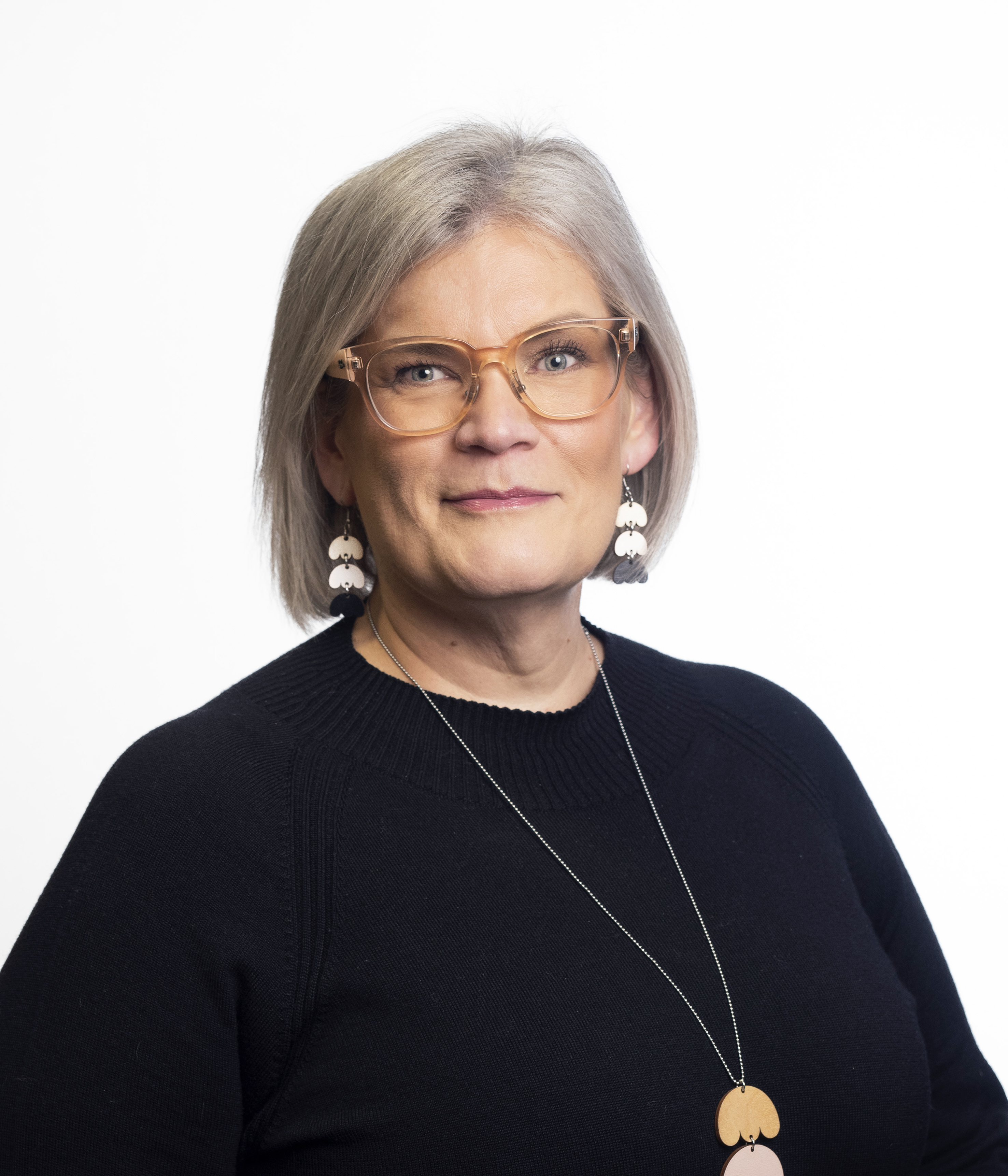}}]{Mirka~Leino} received the Ph.D. degree from Tampere University of Technology, Tampere, Finland, in 2017. She is a Recearching Principal Lecturer and Chief Researcher at RoboAI Research Center with Satakunta University of Applied Sciences, Pori, Finland. Her research interests include machine vision, intelligent and cognitive robotics, AI tools in Automation technologies and technology transfer. 
\end{IEEEbiography}

\vspace{-25pt}

\begin{IEEEbiography}[{\includegraphics[width=1in,height=1.25in,clip,keepaspectratio]{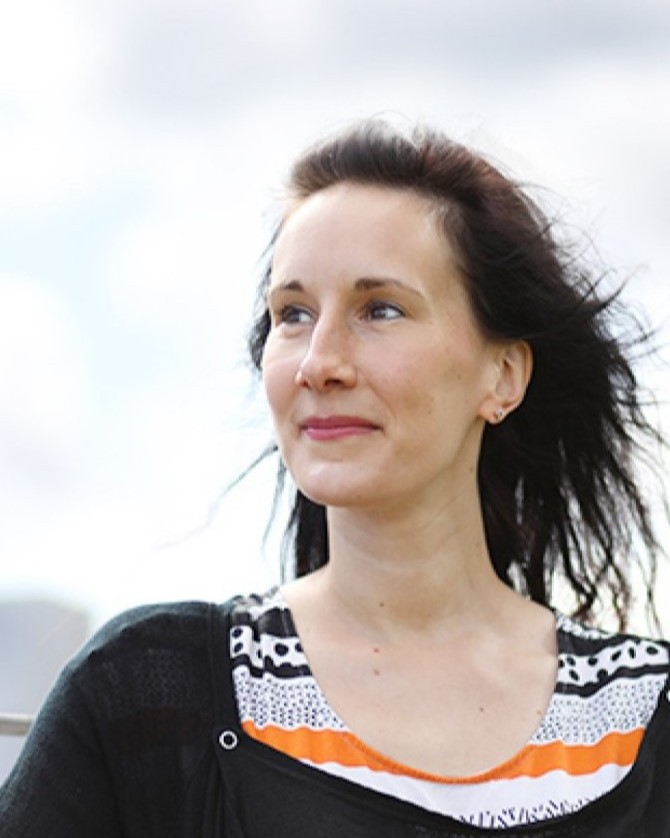}}]{Johanna Virkki}
received the D.Sc. degree in electrical and electronics engineering from Tampere University of Technology, Tampere, Finland, in 2010. She is an Associate Professor with the Faculty of Information Technology and Communication Sciences, Tampere University, Tampere, Finland. Her research interests include augmentative and alternative communication, wearable technology, enabling environments, and gamification of rehabilitation. 
\end{IEEEbiography}

\vfill

\end{document}